\PassOptionsToPackage{unicode}{hyperref}
\PassOptionsToPackage{hyphens}{url}
\PassOptionsToPackage{dvipsnames,svgnames,x11names}{xcolor}
\documentclass[
  12pt]{article}

\usepackage{layouts}
\usepackage[margin=1in]{geometry}
\usepackage{amsmath,amssymb}
\usepackage{iftex}
\ifPDFTeX
  \usepackage[T1]{fontenc}
  \usepackage[utf8]{inputenc}
  \usepackage{textcomp} 
\else 
  \usepackage{unicode-math}
  \defaultfontfeatures{Scale=MatchLowercase}
  \defaultfontfeatures[\rmfamily]{Ligatures=TeX,Scale=1}
\fi
\usepackage{lmodern}
\ifPDFTeX\else  
\fi
\IfFileExists{upquote.sty}{\usepackage{upquote}}{}
\IfFileExists{microtype.sty}{
  \usepackage[]{microtype}
  \UseMicrotypeSet[protrusion]{basicmath} 
}{}
\makeatletter
\@ifundefined{KOMAClassName}{
  \IfFileExists{parskip.sty}{%
    \usepackage{parskip}
  }{
    \setlength{\parindent}{0pt}
    \setlength{\parskip}{6pt plus 2pt minus 1pt}}
}{
  \KOMAoptions{parskip=half}}
\makeatother
\usepackage{xcolor}
\makeatletter
\ifx\paragraph\undefined\else
  \let\oldparagraph\paragraph
  \renewcommand{\paragraph}{
    \@ifstar
      \xxxParagraphStar
      \xxxParagraphNoStar
  }
  \newcommand{\xxxParagraphStar}[1]{\oldparagraph*{#1}\mbox{}}
  \newcommand{\xxxParagraphNoStar}[1]{\oldparagraph{#1}\mbox{}}
\fi
\ifx\subparagraph\undefined\else
  \let\oldsubparagraph\subparagraph
  \renewcommand{\subparagraph}{
    \@ifstar
      \xxxSubParagraphStar
      \xxxSubParagraphNoStar
  }
  \newcommand{\xxxSubParagraphStar}[1]{\oldsubparagraph*{#1}\mbox{}}
  \newcommand{\xxxSubParagraphNoStar}[1]{\oldsubparagraph{#1}\mbox{}}
\fi
\makeatother

\usepackage{longtable,booktabs,array}
\usepackage{calc} 
\usepackage{etoolbox}
\makeatletter
\patchcmd\longtable{\par}{\if@noskipsec\mbox{}\fi\par}{}{}
\makeatother
\IfFileExists{footnotehyper.sty}{\usepackage{footnotehyper}}{\usepackage{footnote}}
\makesavenoteenv{longtable}
\usepackage{graphicx}
\makeatletter
\def\maxwidth{\ifdim\Gin@nat@width>\linewidth\linewidth\else\Gin@nat@width\fi}
\def\maxheight{\ifdim\Gin@nat@height>\textheight\textheight\else\Gin@nat@height\fi}
\makeatother
\setkeys{Gin}{width=\maxwidth,height=\maxheight,keepaspectratio}
\makeatletter
\def\fps@figure{htbp}
\makeatother

\makeatletter
\@ifpackageloaded{caption}{}{\usepackage{caption}}
\AtBeginDocument{%
\ifdefined\contentsname
  \renewcommand*\contentsname{Table of contents}
\else
  \newcommand\contentsname{Table of contents}
\fi
\ifdefined\listfigurename
  \renewcommand*\listfigurename{List of Figures}
\else
  \newcommand\listfigurename{List of Figures}
\fi
\ifdefined\listtablename
  \renewcommand*\listtablename{List of Tables}
\else
  \newcommand\listtablename{List of Tables}
\fi
\ifdefined\figurename
  \renewcommand*\figurename{Figure}
\else
  \newcommand\figurename{Figure}
\fi
\ifdefined\tablename
  \renewcommand*\tablename{Table}
\else
  \newcommand\tablename{Table}
\fi
}
\@ifpackageloaded{float}{}{\usepackage{float}}
\floatstyle{ruled}
\@ifundefined{c@chapter}{\newfloat{codelisting}{h}{lop}}{\newfloat{codelisting}{h}{lop}[chapter]}
\floatname{codelisting}{Listing}

\makeatother
\makeatletter
\@ifpackageloaded{caption}{}{\usepackage{caption}}
\@ifpackageloaded{subcaption}{}{\usepackage{subcaption}}
\makeatother

\ifLuaTeX
  \usepackage{selnolig}  
\fi
\usepackage[]{natbib}
\usepackage{bookmark}

\IfFileExists{xurl.sty}{\usepackage{xurl}}{} 
\hypersetup{
  pdftitle={Title},
  pdfauthor={Author 1; Author 2},
  pdfkeywords={3 to 6 keywords, that do not appear in the title},
  colorlinks=true,
  linkcolor={blue},
  filecolor={Maroon},
  citecolor={Blue},
  urlcolor={Blue},
  pdfcreator={LaTeX via pandoc}}

\newcommand{\anon}{1}

\begin{document}

\def\spacingset#1{\renewcommand{\baselinestretch}%
{#1}\small\normalsize} \spacingset{1}


\if1\anon
{
  \title{\bf A Semiparametric Nonlinear Mixed Effects Model with Penalized Splines Using Automatic Differentiation}
  \author{Matteo D'Alessandro\hspace{.2cm}\\
    Department of Biostatistics, University of Oslo\\
    and \\
    Magne Thoresen\hspace{.2cm}\\
    Department of Biostatistics, University of Oslo\\
    and \\
    Øystein Sørensen\hspace{.2cm}\\
    Department of Psychology, University of Oslo}
  \maketitle
} \fi

\if0\anon
{
  \bigskip
  \bigskip
  \bigskip
  \begin{center}
    {\LARGE\bf A Semiparametric Nonlinear Mixed Effects Model with Penalized Splines Using Automatic Differentiation}
\end{center}
  \medskip
} \fi

\bigskip
\begin{abstract}
We present an estimation procedure for nonlinear mixed-effects models in which the population trajectory is represented by penalized splines and adapted to individuals via subject-specific transformation parameters. By exploiting the mixed model representation of penalized splines, the level of smoothness can be estimated jointly with other variance components. The integration over random effects needed to obtain the marginal likelihood is carried out using the Laplace approximation. Exact derivatives for evaluation and maximization of the resulting likelihood are obtained via automatic differentiation implemented through Template Model Builder. In simulation studies, the method produces improved inferential performance and reduced computational burden when compared to the existing procedures. The approach is further illustrated through a case study on infant height growth in the first two years of life.
\end{abstract}

\noindent%
\vfill

\newpage
\spacingset{1.8} 

\section{Introduction}\label{sec:Introduction}

Longitudinal data, consisting of repeated measurements collected over time on the same experimental units, arise in many areas of research, including medicine, psychology, and the social sciences. Typical examples include growth or developmental trajectories, biomarker profiles such as serum glucose levels following a meal, or repeated assessments of behavioral or cognitive performance over time. Each unit yields a trajectory observed at a finite and possibly irregular set of time points. A common characteristic of such data is that the individual trajectories exhibit a similar underlying shape while differing in scale, timing, or other subject-specific features. Scientific interest often lies in studying both the population-level trajectory and the between-subject variability. By assuming a common trajectory, information can be pooled across subjects, improving estimation even when individual units are observed sparsely.

We consider the case where the population trajectory is estimated semi-parametrically, and adapted to each subject by a small set of individual transformation parameters expressed as random effects. This approach accommodates situations in which an explicit functional form of the population trajectory is not available, by representing it through spline expansions.

Different approaches have been proposed for this data problem.
Self-modeling regression (SEMOR) was first introduced by \citet{Lawton1972} as,
\begin{equation}
f(\boldsymbol{\theta_i}; x) = \theta_{0i} + \theta_{1i}g([x-\theta_{2i}]/\theta_{3i}),
\label{SIM}
\end{equation}
with $ x $ denoting the independent variable and $ f(\boldsymbol{\theta_i}; x) $ the individual-specific trajectory for subject $ i $. The vector $ \boldsymbol{\theta_i} = (\theta_{0i}, \theta_{1i}, \theta_{2i}, \theta_{3i})^{\top} $ contains subject-specific transformation parameters: $ \theta_{0i} $ and $ \theta_{1i} $ correspond to vertical shift and scale, while $ \theta_{2i} $ and $ \theta_{3i} $ control horizontal shift and scale. The function $g$ was modeled as a first-order spline, and estimation of $\boldsymbol{\theta_i}$ and $g$ was achieved via non-linear regression. \citet{Lindstrom1995} extended the model by introducing random effects on the transformation parameters and representing the common shape function via free-knot splines, reducing the number of total parameters. Estimation was achieved through approximate maximum likelihood using the Laplace approximations and the \texttt{nlme} algorithm \citep{Lindstrom1990}, and is now implemented in the package \texttt{sitar} \citep{Cole2010}. Other extensions of SEMOR include those of \citet{Coull2004}, who model multivariate curve data by introducing a latent function shared across curves from the same subject, and \citet{Altman2004}, who extended the model to allow the transformation parameters $\boldsymbol{\theta}$ in (\ref{SIM}) to depend on time-invariant covariates.

\citet{Ke2001} proposed a generalization of SEMOR that places it within a common framework for estimation and inference, alongside many other model classes. The resulting model, known as the semiparametric nonlinear mixed-effects model (SNMM), can be expressed in this form:
\begin{equation}
\begin{aligned}
y_{ij} &= \eta(f, \boldsymbol{\phi}_i, t_{ij}) + e_{ij}, 
 \quad i = 1, \ldots, m, \quad j = 1, \ldots, n_i,\\[4pt]
\boldsymbol{\phi}_i &= \mathbf{A}_i \boldsymbol{\beta} + \mathbf{B}_i \mathbf{b}_i, 
\quad \mathbf{b}_i \sim N(\mathbf{0}, \sigma^2 \mathbf{G}),
\end{aligned}
\label{eq:SNMM}
\end{equation}
where $ y_{ij} $ represents the $ j $th response for subject $ i $, and $ t_{ij} $ is a covariate. The function $ \eta $ is known and depends on $ t_{ij} $, the parameter vector $ \boldsymbol{\phi}_i $ is of dimension $ r $, and $ f$ is an unknown function. The vector $ \boldsymbol{\beta} $ is a $ p $-dimensional vector of fixed effects, while $ \mathbf{b}_i $ is a $ q $-dimensional vector of random effects associated with subject $ i $. $ \mathbf{A}_i $ and $ \mathbf{B}_i $ are design matrices associated with the fixed and random effects. The main advantage of the SNMM is its very general formulation which makes it flexible to various data settings and modeling assumptions. It can also be used to assess the adequacy of a particular functional form in a non-linear mixed model \citep[Section~6]{Ke2001}. The complexity of the SNMM, however, introduces difficulties: the estimation of these types of models is usually done by ``integrating out'' the random effects from the full likelihood to obtain the marginal likelihood; in the case where the random effects enter the likelihood non-linearly, no closed form solutions exist, and approximations are necessary. Moreover, $f$, $\boldsymbol{\beta}$, and $\mathbf{b}$ may interact in ways that complicate estimation.

\citet{Ke2001} model the population curve $f$ with a smoothing spline. The \textit{double-penalized marginal likelihood} (which includes a penalty related to smoothness and one to the subject random effects) is optimized iteratively by use of \texttt{nlme} and a routine for spline estimation. The method is implemented in the R package \texttt{assist} \citep{Wang2004}, and procedures for inference on both the parametric components and the nonparametric function are provided.

This approach has some limitations: (i) as noted by \citet{Elmi2011}, the estimation algorithm separates the likelihood associated with the shape function from that associated with the fixed and random effects, which does not guarantee that the iterative procedure converges to a maximizer of the joint likelihood. Moreover, variance estimates obtained from separate likelihoods will likely fail to fully account for the uncertainty, leading to unsatisfactory inference; (ii) the use of a full-rank smoothing spline, which is the default configuration in \texttt{assist}, implies a basis dimension equal to the number of observed data points, which can lead to substantial computational burden; in many applications, comparable fits can be achieved using lower-rank spline bases with far fewer basis functions \citep[Chapter~5.2]{Wood2017}; (iii) because the mixed model representation of penalized splines is not exploited, the smoothing parameter must be selected separately from the remaining model components in a data-adaptive manner at each iteration. This further increases computational cost and limits the scalability of the method to large datasets.

Other estimation procedures for (\ref{eq:SNMM}) have been considered in the literature. \citet{Elmi2011} represent the population curve using B-splines and evaluate the approximate marginal likelihood via adaptive Gaussian quadrature. Although treating the spline coefficients as fixed effects reduces the dimensionality of the integral, the computational cost of Gaussian quadrature still increases rapidly with the dimension of the random effects. In addition, because smoothness is controlled solely through the number of knots rather than a penalty, this user choice becomes critical for obtaining an appropriate level of smoothness. More recent work \citep{ArribasGil2013, DelaCruz2017, Mrquez2023} consider additional semi-parametric approaches (e.g. a LASSO approach on a set of basis functions), but focus on the case where the non-linear function does not depend on random effects, so that they can be integrated out explicitly without need for approximation. This greatly simplifies the method, but restricts its capabilities.

A closely related framework is that of generalized additive latent and mixed models (GALAMM, \citet{sorensenLongitudinalModelingAgeDependent2023}), which similarly combines smooth functions and latent variables within a single model, and is estimated using the Laplace approximation and automatic differentiation.  In the GALAMM formulation, however, smooth terms are only included as functions of an observed covariate, with no random effects appearing inside their argument. This precludes fitting the SNMM in (\ref{eq:SNMM}), where the transformation $\gamma(\phi_i, t)$ can depend on subject-specific random effects. In conclusion, a fully satisfactory algorithm for SNMMs is not yet available, which has restricted their widespread use.

In this paper, we develop an estimation procedure for the SNMM based on a mixed-model representation of P-splines, which enables joint estimation of smoothness through variance components. The marginal likelihood is obtained via the Laplace approximation, with the required derivatives computed using automatic differentiation (AD), yielding efficient and accurate evaluation. The method is implemented using the R package \texttt{TMB} \citep{Kristensen2015}. The use of a unified likelihood, together with the availability of higher-order derivatives, allows inference on the population curve and the model parameters in a straightforward way directly from \texttt{TMB}. More generally, this approach gives us access to the usual likelihood machinery for quantifying uncertainty, testing hypotheses, and comparing different model specifications.

The article is structured as follows. Section \ref{sec:SNMM} describes our method in detail, shows the derivation of the marginal likelihood for the model through the Laplace approximation, and how AD is used to compute the necessary derivatives for estimation and inference; Section \ref{sec:Simulation} presents simulated examples to study the performance of the method in comparison to the \texttt{assist} package, and the adaptive Gaussian Quadrature method of \citet{Elmi2011}; in Section \ref{sec:Application}, the method is illustrated using a dataset of height measurements of infants from birth to two years of age and compared to \texttt{sitar}, and we show the use of a parametric bootstrap procedure to check the validity of the Laplace approximation in applications; a discussion of the results and final remarks are included in Section \ref{sec:Discussion}.

\section{Methods}\label{sec:SNMM}

\subsection{Estimation}

We here describe the proposed estimation procedure for model \eqref{eq:SNMM}. In our notation, the dependence of the argument of $f$ on the other parameters is made explicit by writing  $f(\gamma(\boldsymbol{\phi}_i; t_{ij}))$, where $\gamma$ is the x-axis transformation that $f$ is subject to. We also define $u_i = \gamma(\boldsymbol{\phi}_i; t)$, such that $f$ is a function of $u_i$.

Assume $f$ is constructed as a weighted sum,
\begin{equation*}
    f(u) = \sum_{k=1}^{K} \zeta_k c_k(u),
\end{equation*}
where $c_k(\cdot)$ is the $k$-th basis function and 
$\boldsymbol{\zeta} = (\zeta_1, \ldots, \zeta_K)^\top$ are the corresponding spline coefficients. We can write 
\begin{equation}
    f(\gamma(\boldsymbol{\phi}; t)) = \mathbf{X}(\boldsymbol{\phi}, t) \boldsymbol{\zeta},
    \label{spline_design_matrix}
\end{equation}
where the design matrix $\mathbf{X}$ has elements $X_{ij} = c_j(\gamma(\boldsymbol{\phi_i}; t_{i}))$. Note that the design matrix in our model depends on the parameters $\boldsymbol{\beta}$ and $\mathbf{b}$ through $\boldsymbol{\phi}$. 

The level of smoothness is tuned by penalizing wiggliness, namely by a penalty of the form $\lambda \boldsymbol{\zeta}^\top \mathbf{S} \boldsymbol{\zeta}$, where $\mathbf{S}$ is a penalty matrix (for example, based on finite differences of adjacent coefficients), and $\lambda > 0$ is a smoothing parameter. Throughout, we take $c_k(\cdot)$ to be B-spline basis functions of a fixed degree, so that, together with the penalty on $\boldsymbol{\zeta}$, $f$ corresponds to a P-spline \citep{Eilers1996}.

It is known that smoothing splines can be represented in a mixed model form \citep[among others]{Kimeldorf1970,Wang1998,Lin1999}, in which components of the smooth in the null space of the penalty are treated as fixed effects, and the penalized components as random effects. This allows smoothing parameters to be estimated via restricted maximum likelihood (REML) \citep{Patterson1971}. Following \citet{Wood2004}, we can form the eigendecomposition $\mathbf{S} = \mathbf{U}\mathbf{D}\mathbf{U}^\top$, where $\mathbf{U}$ is orthonormal and $\mathbf{D}$ is diagonal (with eigenvalues arranged in decreasing order along the leading diagonal). We form $\boldsymbol{\zeta}_u = \mathbf{U}^{\top} \boldsymbol{\zeta}$. Since $\mathbf{S}$ is generally rank-deficient, the last few elements on the leading diagonal of $\mathbf{D}$ will be zero. Let $\mathbf{D}^+$ denote the submatrix of $\mathbf{D}$ containing only the nonzero elements on the leading diagonal, of dimension $r \times r$. Partition $\boldsymbol{\zeta}_u$ so that
$\boldsymbol{\zeta}_u^{\top} = [\boldsymbol{\omega}_u^{\top}, \boldsymbol{\zeta}_F^{\top}]$ and $\boldsymbol{\zeta}_u^{\top} \mathbf{D} \boldsymbol{\zeta}_u = \boldsymbol{\omega}_u^{\top} \mathbf{D}^+ \boldsymbol{\omega}_u$.
Similarly, partition $\mathbf{X}\mathbf{U}$ by letting $\mathbf{X}_F$ contain the columns corresponding to zero entries on the diagonal of $\mathbf{D}$, and  $\mathbf{X}_u$ the ones corresponding to $\mathbf{D^+}$.
By letting $\boldsymbol{\omega} = \sqrt{\mathbf{D}^+}\boldsymbol{\omega}_u$, the penalty $\lambda \boldsymbol{\zeta}^\top \mathbf{S} \boldsymbol{\zeta}$ becomes $\lambda \boldsymbol{\omega}^\top\boldsymbol{\omega}$, which is equivalent to adding random effects $\boldsymbol{\omega} \sim N(0,\frac{1}{\lambda} \mathbf{I)}$. Defining $\mathbf{X}_R = \mathbf{X}_u (\sqrt{\mathbf{D}^{+}})^{-1}$, the matrix representation in (\ref{spline_design_matrix}) can be reparametrized as 
\begin{equation*}
f =\mathbf{X}_F \boldsymbol{\zeta}_F + \mathbf{X}_R \boldsymbol{\omega}.
\end{equation*}

Now we go on to obtain the marginal likelihood for the model. Let $\boldsymbol{\psi} = (\mathbf{b},\boldsymbol{\omega})$ be the complete vector of random effects, with $\mathbf{b}_{i}$ associated with the subjects, and $\boldsymbol{\omega}$ coming from the representation of the smooth. Let $\boldsymbol{\theta}= (\boldsymbol{\theta}_F, \boldsymbol{\beta}, \boldsymbol{\Omega})$ be the vector of all fixed effects, which includes the unpenalized spline coefficients $\boldsymbol{\theta}_F$, the fixed effects $\boldsymbol{\beta}$ and $\boldsymbol{\Omega}$ collecting the variance parameters, which determine $\boldsymbol{\sigma}$ and the random-effects covariance matrix $\boldsymbol{G}$. Finally, let $\boldsymbol{\widetilde{G}} = \mathbf{I}_m \otimes \boldsymbol{G}$ denote the block-diagonal covariance matrix of the random effects across all subjects. We can establish the conditional density of $\boldsymbol{y}$ given $\boldsymbol{\psi}$ as,
\begin{equation*}
    p(\boldsymbol{y}|\boldsymbol{\psi}) = (2\pi\sigma^2)^{-\frac{N}{2}} \text{exp}\left\{-\frac{1}{2\sigma^2}\left[y - \eta(\phi, f)\right]^2\right\},
\end{equation*}
and the density of the random effects $\boldsymbol{\psi}$,
\begin{equation*}
\begin{aligned}
p(\boldsymbol{\psi})
&= p(\mathbf{b})p(\boldsymbol{\omega}) \\
&= (2\pi\sigma^2)^{-mq/2}|\boldsymbol{\widetilde{G}}|^{-1/2}\,
  (2\pi)^{-r/2}\lambda^{r/2}
  \exp\!\left(
    -\frac{1}{2\sigma^2}\mathbf{b}^\top\boldsymbol{\widetilde{G}}^{-1}\mathbf{b}
    -\frac{\lambda}{2}\|\boldsymbol{\omega}\|^2
  \right),
    \end{aligned}
\end{equation*}
which gives the joint density
\begin{equation}
    \begin{aligned}
p(\mathbf{y}, \boldsymbol{\psi})
&= p(\mathbf{y}\mid\boldsymbol{\psi})\,p(\boldsymbol{\psi}) \\[4pt]
&= (2\pi)^{-(N+mq+r)/2}\,
   (\sigma^2)^{-(N+mq)/2}\,
   |\boldsymbol{\widetilde{G}}|^{-1/2}\,
   \lambda^{r/2} \\[4pt]
&\qquad\times
   \exp\!\left(
     -\frac{1}{2\sigma^2}\|\mathbf{y}-\eta(\phi,f)\|^2
     -\frac{1}{2\sigma^2}\mathbf{b}^\top\boldsymbol{\widetilde{G}}^{-1}\mathbf{b}
     -\frac{\lambda}{2}\|\boldsymbol{\omega}\|^2
   \right).
\end{aligned}
    \label{joint_density}
\end{equation}

The fixed parameters can be estimated by maximizing the marginal likelihood, obtained by integrating out the random effects from the joint density:
\begin{equation}
     \begin{aligned}
L(\boldsymbol{\theta}) 
&=\; \int p(\mathbf{y}\mid\boldsymbol{\psi})\,p(\boldsymbol\psi)\;
   \mathrm{d}\boldsymbol\psi \\[8pt]
&=  (2\pi)^{-(N+mq+r)/2}\,
   (\sigma^2)^{-(N+mq)/2}\,
   |\boldsymbol{\widetilde{G}}|^{-1/2}\,
   \lambda^{r/2} \int \exp\!\left\{ g(\boldsymbol{\psi})\right\}\,
\mathrm{d}\boldsymbol\psi, \; 
\end{aligned}
    \label{marginal_likelihood}
\end{equation}
with the exponent 
\begin{equation*}
    g(\boldsymbol{\psi}) = -\frac{1}{2\sigma^2}\|\mathbf{y}-\eta(\phi,f)\|^2
     -\frac{1}{2\sigma^2}\mathbf{b}^\top\boldsymbol{\widetilde{G}}^{-1}\mathbf{b}
     -\frac{\lambda}{2}\|\boldsymbol{\omega}\|^2
\end{equation*}
from (\ref{joint_density}). 

The $(mq+r)$-dimensional integral in (\ref{marginal_likelihood}) is generally intractable, because $\eta$ depends non-linearly on the random effects. The Laplace approximation is often used to approximate integrals in this setting \citep{Ke2001,Wood2010}, as it replaces the original integral with a closed-form expression based on a local Gaussian approximation to the integrand around its mode
$\hat{\boldsymbol{\psi}} = \arg\max_{\boldsymbol{\psi}} g(\boldsymbol{\psi})$. This yields
\begin{equation*}
g(\boldsymbol{\psi}) \approx g(\hat{\boldsymbol{\psi}}) - \tfrac{1}{2}
(\boldsymbol{\psi} - \hat{\boldsymbol{\psi}})^{\top}
\mathbf{H}(\hat{\boldsymbol{\psi}})
(\boldsymbol{\psi} - \hat{\boldsymbol{\psi}}),
\end{equation*}
where $\mathbf{H}(\hat{\boldsymbol{\psi}}) = -\nabla^2 g(\hat{\boldsymbol{\psi}})$ is the negative Hessian matrix of $g$ evaluated at the mode. It follows that the Laplace approximate marginal log-likelihood is
\begin{equation}
    \begin{aligned}
    l(\boldsymbol{\theta}) =& -\frac{N}{2}\log(2\pi) - (N+mq) \log(\sigma) - \frac{1}{2}\log\left|\boldsymbol{\widetilde{G}}\right| + \frac{r}{2}\lambda\\
   &- \frac{1}{2}\log \left|\mathbf{H}(\boldsymbol{\theta})\right| + g\left\{\hat{\boldsymbol{\psi}}(\boldsymbol{\theta}), \boldsymbol{\theta}\right\}   ,
    \end{aligned}
    \label{marginal_log_likelihood}
\end{equation}
where we have expressed explicitly the dependence of $\mathbf{H}$, $\hat{\boldsymbol{\psi}}$, and $g$ on $\boldsymbol{\theta}$.

Computation of the Laplace approximation involves two main steps: (1) obtaining the mode by maximizing $g(\boldsymbol{\psi})$, and (2) evaluating the Hessian of $g$ at that mode. Both steps require first and second-order derivatives of $g$ with respect to $\boldsymbol{\psi}$. Given the complexity of the model, deriving these expressions by hand would be tedious, error-prone, and their final form would still depend on the specific functional choice of $\eta(\cdot)$, and the basis functions used. To avoid these difficulties, AD was employed to compute the required derivatives of $g$. AD computes numerical derivatives accurate to machine precision by systematically applying the chain rule to every elementary operation in a program \citep{Griewank2008, Skaug2006}.
Our implementation used the \texttt{R} package Template Model Builder, \texttt{TMB} \citep{Kristensen2015}, which relies on \texttt{CppAD} for forward and reverse mode AD via operator overloading \citep{CppADmanual}.

The estimation in \texttt{TMB} proceeds hierarchically. For starting values of the fixed parameters, $\boldsymbol{\theta}$, \texttt{TMB} first locates the conditional mode of the random effects, $\hat{\boldsymbol{\psi}}(\boldsymbol{\theta}) = \arg\max_{\boldsymbol{\psi}} g(\boldsymbol{\psi}; \boldsymbol{\theta})$, using Newton optimization. At the optimum, the Hessian with respect to the random effects is computed for the Laplace approximation to the marginal likelihood. \texttt{TMB} then computes the gradient of this Laplace-approximated objective with respect to $\boldsymbol{\theta}$. This allows the use of any gradient-based optimization routine to find the maximum of the marginal likelihood with respect to the fixed parameters: in our implementation we used the \texttt{nlminb} function in \texttt{R}, which employs a quasi-Newton trust-region method from the PORT routines \citep{Gay1990}. The final estimates of the fixed effects correspond to the maximizer of the marginal likelihood, while the estimates of the random effects are given by their modes evaluated at the optimal values of the fixed parameters.

\subsection{Inference}\label{sec: Inference}

At convergence of the optimization problem, the covariance matrix of the estimated fixed effects $\mathrm{Var}(\hat{\boldsymbol{\theta}})$ can be obtained from the inverse of the observed Hessian of the (approximated) marginal log-likelihood (\ref{marginal_log_likelihood}),
\begin{equation*}
\mathrm{Var}(\hat{\boldsymbol{\theta}}) 
= \left[-\nabla^2_{\boldsymbol{\theta}} l(\hat{\boldsymbol{\theta}})\right]^{-1}.
\end{equation*}
The delta method is then used by \texttt{TMB} to obtain a first-order approximation to the covariance matrix of any differentiable function of the fixed parameters $\phi(\hat{\boldsymbol{\theta}})$:
\begin{equation*}
\mathrm{Var}(\phi(\hat{\boldsymbol{\theta}})) 
\approx 
\mathbf{J}_{\phi}\,
\mathrm{Var}(\hat{\boldsymbol{\theta}})\,
\mathbf{J}_{\phi}^{\top},
\qquad 
\mathbf{J}_{\phi}
=
\left.
\frac{\partial \phi(\boldsymbol{\theta})}{\partial \boldsymbol{\theta}^{\top}}
\right|_{\boldsymbol{\theta}=\hat{\boldsymbol{\theta}}}.
\end{equation*}
In order to obtain inferences for the population and subject-level spline curves, the method has to be extended to functions of both fixed and random effects. However, since the estimate of the random effects $\hat{\boldsymbol{\psi}}(\boldsymbol{\theta})$ depends on $\boldsymbol{\theta}$, its variance has to take into account the uncertainty in the estimation of $\boldsymbol{\theta}$, which requires further derivation.

We are interested in the covariance matrix of the estimation errors on the random and fixed effects, which is given by
\begin{equation*}
\mathrm{Var}
\begin{pmatrix}
\widehat{\boldsymbol{\psi}} - \boldsymbol{\psi} \\
\widehat{\boldsymbol{\theta}} - \boldsymbol{\theta}
\end{pmatrix}
\approx
\begin{pmatrix}
\mathbf{H}^{-1} & \mathbf{0} \\
\mathbf{0} & \mathbf{0}
\end{pmatrix}
+
\mathbf{J}\,
\mathrm{Var}(\widehat{\boldsymbol{\theta}})\,
\mathbf{J}^{\top},
\end{equation*}

where $\mathbf{J}$ is the Jacobian of $[\widehat{\boldsymbol{\psi}}(\boldsymbol{\theta}), \boldsymbol{\theta}]^{T}$ with respect to~$\boldsymbol{\theta}$ (the derivation of this formula can be found in the book version of \citet{TMBdocum}, Appendix 16.3). With this, we can use the delta method to express the variance of the estimation error on a general function of fixed and random effects $f(\boldsymbol{\theta}, \boldsymbol{\psi})$:
\begin{equation*}
\mathrm{Var}
\!\left(
f(\widehat{\boldsymbol{\psi}}, \widehat{\boldsymbol{\theta}})
- 
f(\boldsymbol{\psi}, \boldsymbol{\theta})
\right)
\approx
\nabla{f}\,
\mathrm{Var}
\begin{pmatrix}
\widehat{\boldsymbol{\psi}} - \boldsymbol{\psi} \\[3pt]
\widehat{\boldsymbol{\theta}} - \boldsymbol{\theta}
\end{pmatrix}
\nabla{f}^{\top},
\end{equation*}
where $\nabla f = \left(\partial f/\partial \boldsymbol{\psi},\ \partial f/\partial \boldsymbol{\theta}\right)$ denotes the gradient of $f$ with respect to the full vector obtained by stacking $\boldsymbol{\psi}$ and $\boldsymbol{\theta}$. \citet{Zheng2021} discuss in detail the interpretation of this variance, which they denote \textit{prediction variance}. When using prediction variance, inferences about the random effects will be appropriate only for the specific values of the random effects of the observed data. This procedure, however, does not describe the variability of the estimator $\hat{\boldsymbol{\psi}}$ caused by the repeated sampling of $\boldsymbol{\psi}$.

To construct a confidence band, we first extract from \texttt{TMB} the estimated values of the target function evaluated on a grid of points, denoted by $\hat{\mathbf{f}}$, together with the corresponding covariance matrix of the estimation errors, $\hat{\mathbf{V}}$. A pointwise Wald-type confidence band with nominal level $(1-\alpha)100\%$ is then given by $
\hat{\mathbf{f}} \pm z_{1-\alpha/2}\,\sqrt{\operatorname{diag}(\hat{\mathbf{V}})}$,
where $z_{1-\alpha/2}$ denotes the $1-\alpha/2$ quantile of the standard normal distribution.
Confidence bands constructed in this manner attain coverage close to the nominal level on average over the domain of the function. In contrast, obtaining simultaneous coverage of the entire function with probability $(1-\alpha)100\%$ requires replacing the normal critical value with one obtained via simulation. Further methodological details are provided in \cite[Section~1.1]{sorensenLongitudinalModelingAgeDependent2023}.

\subsection{Further considerations}

\subsubsection{Knot selection}\label{sec: Knot Selection}

Because $f$ is a function of the latent values $\gamma(\boldsymbol{\phi}; t)$, the appropriate locations for knots are not directly observable. Moreover, knot placement determines the structure of the penalty matrix~$\mathbf{S}$, which must be decomposed into its null space and range space to express the spline coefficients as fixed and random effects. Allowing the knot locations to depend on the random effects would cause the matrix~$\mathbf{S}$ and its eigendecomposition to depend on the model parameters, and incorporating such a procedure within AD would be infeasible. For these reasons, we would prefer fixing the knot locations throughout estimation.

We consider two approaches to do this, depending on the nature of $\gamma$. The first applies when $\gamma$ is known to take values in a finite and fixed interval, as in the first simulation study presented in Section~\ref{sec:Simulation}. In this case, the range of $\gamma$ is known, and the knots can therefore be fixed on a given interval which safely covers the support of the function.

In the second case, we consider functions whose domains are not bounded a priori. To retain a fixed set of knots, we modify $\gamma$ so that its codomain is predictable over the observed range of $t$ and over plausible values of the random effects. Specifically, the lower and upper bounds of $\gamma$ are defined using the observed values of $t$ and values of $\boldsymbol{\phi}$ within $\pm 3$ standard deviations of their current estimated variances. To preserve compatibility with AD, these bounds are expressed as smooth functions of the model parameters.

For example, if $\gamma(\boldsymbol{\phi}; t) = t - b_2$ (as in the second simulation of Section~\ref{sec:Simulation}), the bounds can be defined as
\begin{equation*}
\min \gamma = \min(t) - 3\,\sigma_{b_2},
\qquad
\max \gamma = \max(t) + 3\,\sigma_{b_2},
\end{equation*}
so that both depend smoothly on the estimated variance of $b_2$. The transformed variable is then standardized as
\begin{equation*}
\gamma^*(\boldsymbol{\phi}, t, \boldsymbol{\Omega}) 
= \frac{\gamma(\boldsymbol{\phi}; t) - \min \gamma(\boldsymbol{\Omega})}
{\max \gamma(\boldsymbol{\Omega}) - \min \gamma(\boldsymbol{\Omega})},
\qquad \gamma^*(\boldsymbol{\phi}; t) \in [0,1].
\end{equation*}
By construction, $\gamma^*$ takes values in $[0,1]$ over the relevant domain of the variables. The knot locations are selected once on this fixed interval before optimization and not moved during estimation. As the variance parameters $\boldsymbol{\Omega}$ are updated, only the standardization $\gamma^*(\phi,t,\boldsymbol{\Omega})$ changes through its dependence on $\sigma_{b_3}$, thereby remapping the transformed observations to the same fixed set of knots. On this standardized scale the spline basis and the penalty matrix therefore remain unchanged throughout optimization, while the transformation varies smoothly with $\boldsymbol{\Omega}$, preserving differentiability.
However, on the scale on which $f$ is interpreted, the effective knot positions depend on $\boldsymbol{\Omega}$, and two values of $\boldsymbol{\Omega}$ correspond to two different spline spaces, so that the resulting objective is not equivalent to the
marginal likelihood of a model with a fixed basis.

\subsubsection{Starting values}\label{sec:starting_values}

Estimation of the approximated marginal likelihood is performed using a gradient-based optimization procedure and therefore requires starting values for all model parameters; however, finding suitable values is often nontrivial.

Following \citet{Elmi2011}, we suggest adopting a two-step strategy. First, starting values for the spline coefficients can be obtained by fitting a generalized additive model to the raw data. Second, treating this fitted curve as known, a nonlinear mixed-effects model can be fitted using the R package \texttt{nlme} \citep{Pinheiro2000} to obtain starting values for the fixed effects and random-effects variance components. Although this procedure does not involve joint estimation, it is computationally efficient and has shown to provide sensible starting values in our experience.

\section{Simulation studies}\label{sec:Simulation}

First, we compare our estimation and inference procedure (labeled ``snmmTMB'') with the R package \texttt{assist} and with the adaptive Gauss--Hermite quadrature procedure of \citet{Elmi2011}, which we adapt from the available supplementary code with the original paper and label ``snmmAGQ''. We consider the same simulated data setup as reported in \citet{Ke2001}. Specifically, we generate data from the model
\begin{equation*}
y_{ij}
= 5 + b_{1i}
+ 2\exp(b_{2i}) \sin \left[ 2 \pi \!\left( t_{j} - \frac{\exp(b_{3i})}{1+\exp(b_{3i})} \right)\right] + \varepsilon_{ij},
\quad i = 1, \ldots, n,\; j = 1, \ldots, m,
\end{equation*}
with $t_{j} = (j-1)/(m-1) \in [0,1]$, errors $\varepsilon_{ij} \sim\mathcal{N}(0, \sigma^2)$ and random effects vector $\mathbf{b}_i = (b_{1i}, b_{2i}, b_{3i})^{\top} \sim \mathcal{N}(\mathbf{0}, \sigma^2\mathbf{D})$. In this formulation, $b_{1i}$ represents a subject vertical shift, $b_{2i}$ is a curve scaling factor, and $b_{3i}$ introduces a subject phase shift.

\begin{figure}
\vspace{-2cm}
\centering
    \includegraphics[width=\linewidth]{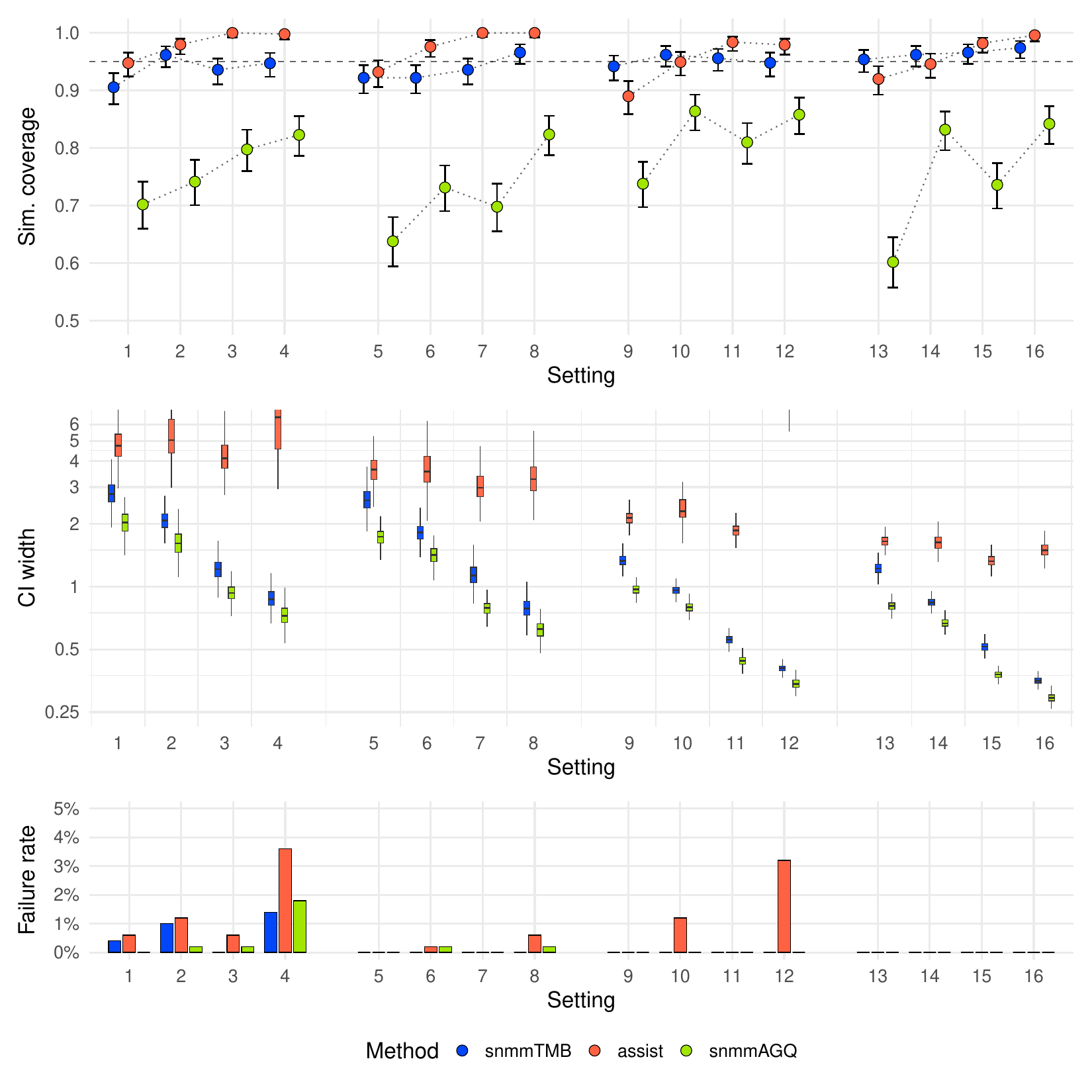}
    \caption{\textit{Results of the sine curve simulation on the population curve}. The 16 settings are indexed so that each group of $4$ settings has the same sample size $(n, m)$. The first four settings correspond to $(n, m, \sigma, \boldsymbol{D})$ equal to $(10,10,1,\text{high})$, $(10,10,1,\text{low})$, $(10,10,0.4,\text{high})$, and $(10,10,0.4,\text{low})$, with the remaining groups of 4 following the same order but with $(n, m)$ equal to $(10,20)$, $(50,10)$, and $(50,20)$. The top panel displays the average simultaneous coverage of the population curve across 500 replications, with bars indicating 95\% Clopper--Pearson confidence intervals for the coverage probability. The middle panel shows boxplots of the confidence-band width, on a logarithmic scale; the box for \texttt{assist} in setting 12, with median $15.71$, falls outside the plotting range. The bottom panel shows the percentage of replications that failed to produce a fit.}
    \label{fig:sinecurve_population}
\end{figure}

\begin{figure}
\centering
    \includegraphics[width=\linewidth]{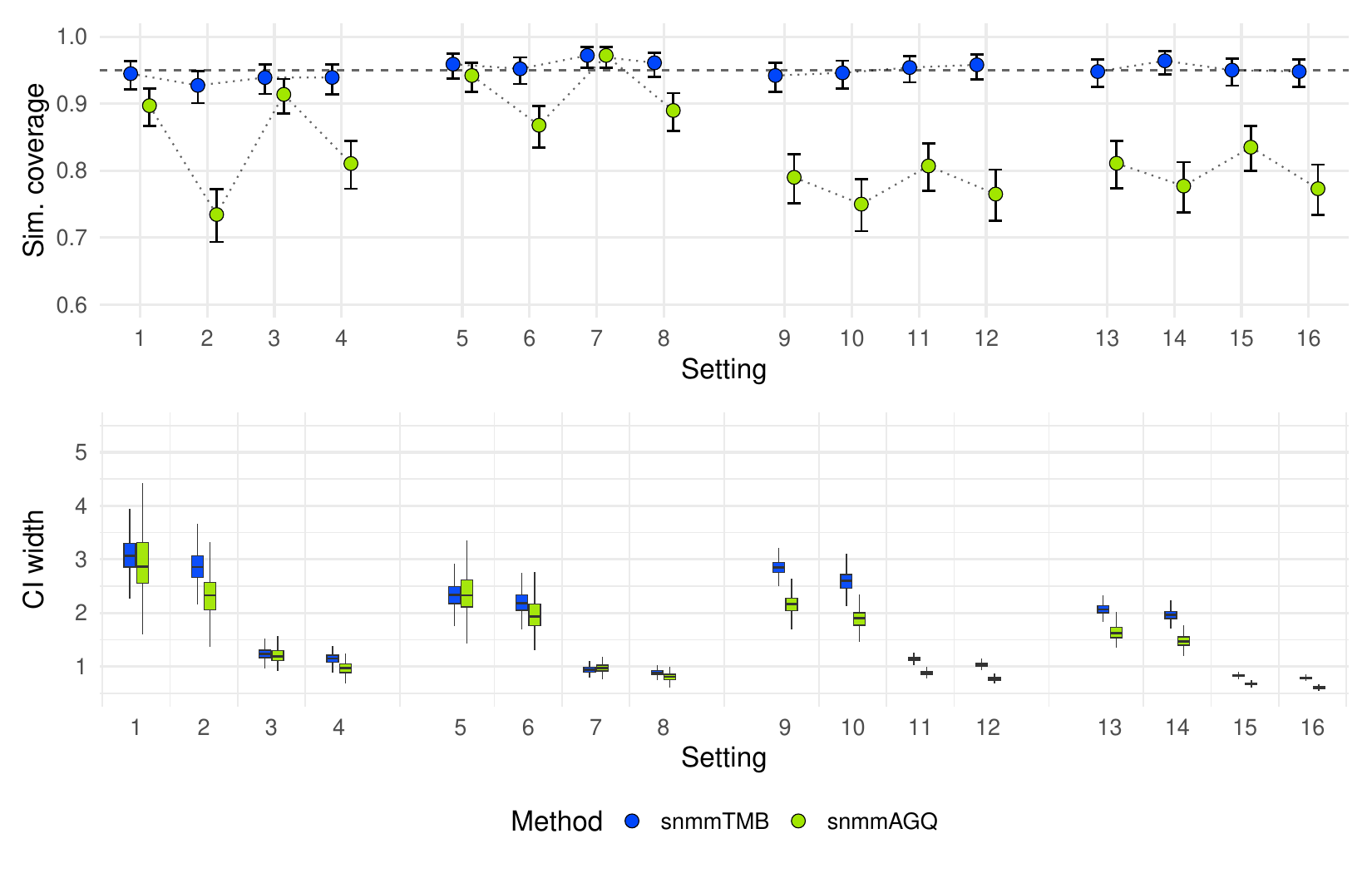}
    \caption{\textit{Results of the sine curve simulation on the subject curves}. The parameter settings follow those of Figure \ref{fig:sinecurve_population}. The top panel displays the average simultaneous coverage of the subject curves across 500 replications, with bars indicating 95\% Clopper--Pearson intervals for the coverage probability. The bottom panel shows boxplots of the confidence-band width. \texttt{assist} does not produce subject-level curves and is therefore not shown.}
    \label{fig:sinecurve_subject}
\end{figure}

We consider all possible combinations of parameter values on the grid: (1) $\sigma \in \{1, 0.4\}$; (2) $m \in \{10, 20\}$; (3) $n \in \{10, 50\}$; (4) low-variance $D = \mathrm{diag}(0.25, 0.16, 0.04)$, or high-variance $D = \mathrm{diag}(1, 0.25, 0.16)$; for a total of 16 data settings. Each setting is simulated 500 times to obtain the results of Figures~\ref{fig:sinecurve_population} and~\ref{fig:sinecurve_subject}.

For snmmTMB, we use a cubic spline basis with $K = 15$ basis functions with a penalty based on second-order finite differences and subject to a sum-to-zero constraint. The knots are distributed uniformly in $[-1, 1]$, as the phase shift function $\gamma(\mathbf{b}, t_{ij}) = t_{ij} - \exp(b_{3i})/(1+\exp(b_{3i}))$ takes values in this interval, independently of the values of $b_{3i}$, and they are held fixed throughout estimation. Starting values are obtained from an \texttt{nlme} fit approximating the model.

The \texttt{assist} method specifications are chosen in line with the description of the simulation in \citet{Ke2001}; we need to provide a null space for $f$, which is chosen as the periodic functions $\mathrm{span}\{\sin(2\pi x), \cos(2\pi x)\}$, together with the corresponding reproducing kernel. Note that the model is thus given information about the periodicity of the underlying population curve. Since the package does not report simultaneous bands, we construct them from the fitted object as described in Online Appendix A. Convergence tolerances, iteration limits, and smoothing parameter selection are left at their package defaults. The exact calls are given in the code supplement.

The snmmAGQ implementation integrates the subject random effects by adaptive Gauss--Hermite quadrature with $5$ nodes per random-effect dimension. Its spline trajectory has no smoothing penalty, so choosing the number of knots is necessary to obtain the correct level of smoothness; following \citet{Elmi2011}, the number of basis functions is selected by AIC in the range from $4$ to $10$ for every simulated dataset. Starting values are obtained by the same two-step procedure of Section \ref{sec:starting_values}. We implemented standard error calculations from Equation~(2.16) of \citet{Elmi2011} with the numerical Hessian of the log-likelihood, and obtained simultaneous confidence bands by the same simulation used for snmmTMB.

In Figure~\ref{fig:sinecurve_population} we compare: (i) the simultaneous coverage of the population curve, (ii) the width of the confidence band and (iii) the percentage of iterations that failed to produce a fit. Both coverage and confidence band width averages are computed over the successful replications only. We can see that coverage for snmmTMB is stable around the nominal value in all settings (0.906--0.974, mean 0.948). \texttt{assist} also achieves good coverage in most settings; its bands, however, are consistently wider, and the method exhibits a tendency to over-cover in the low-noise settings. snmmAGQ shows the opposite pattern, undercovering in every setting, with bands comparable to or narrower than snmmTMB's. The overall average number of basis functions selected by AIC among all settings is $6.7$.

We also compare the confidence bands for subject-level curves, which combine the uncertainty of the population curve and the subject random effects. Figure~\ref{fig:sinecurve_subject} shows that snmmTMB attains close to nominal coverage in this task as well, while snmmAGQ again undercovers; \texttt{assist} does not produce subject-level bands and is therefore absent from this comparison. As expected, the confidence intervals are much larger in the settings with fewer observations per subject, as there is more uncertainty on the subject curve.

\begin{figure}
\centering
    \includegraphics[width=\linewidth]{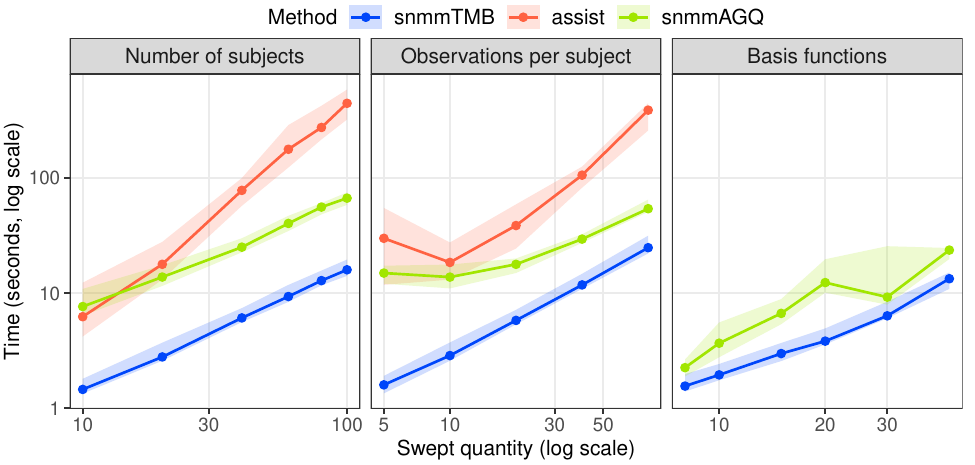}
    \caption{\textit{Computational time and scalability on the sine curve simulation}. Data are generated with $\sigma = 1$ and $\mathbf{D} = \operatorname{diag}(1, 0.25, 0.16)$. Each panel varies one quantity around the baseline configuration $n = 20$, $m = 10$, $K = 15$: the number of subjects $n \in \{10, 20, 40, 60, 80, 100\}$ (left), the number of observations per subject $m \in \{5, 10, 20, 40, 80\}$ (middle), and the number of basis functions $K \in \{8, 10, 15, 20, 30, 45\}$ (right). Lines are medians over 50 replications and ribbons are interquartile ranges. \texttt{assist} does not appear in the right panel, as it places a knot at every distinct design point and therefore has no basis dimension to vary.}
    \label{fig:sinecurve_scalability}
\end{figure}

Finally, Figure~\ref{fig:sinecurve_scalability} presents results on computational time and scalability. We consider a baseline configuration of $n = 20$ subjects, $m = 10$ observations per subject and $K = 15$ basis functions, and vary one quantity at a time: $n$ up to $100$, $m $ up to $80$ and $K$ up to $45$. For snmmAGQ, the number of knots is selected by AIC as in the simulation study when varying $n$ and $m$, and its reported time includes the selection; when varying the basis size, both methods use the same fixed $K$. We observe that snmmTMB is substantially faster than \texttt{assist} throughout, with a wider gap at large sample sizes: at $n = 100$ the median fitting time is around 16 seconds for snmmTMB against 450 seconds for \texttt{assist}. snmmAGQ, whose time includes multiple fittings for the AIC knot selection, is comparable to \texttt{assist} at small sample sizes but grows at a rate similar to snmmTMB, which is close to linear in the number of subjects and in the number of observations per subject. As discussed in \citet[Section~4]{Kristensen2015}, the cost of a TMB evaluation is dominated by the Laplace approximation, which is largely determined by the sparsity of the Hessian of the joint negative log-likelihood with respect to the random effects. In our simulations, this matrix is bordered block-diagonal, with one block per subject and a border corresponding to the spline coefficients. The number of non-zero entries therefore grows linearly with the number of subjects, and only mildly with the number of basis functions, which is why Figure~\ref{fig:sinecurve_scalability} shows a large basis to be affordable, with the penalty left to control smoothness. Increasing the number of random effects per subject $q$ would also only enlarge the diagonal blocks, so the cost of the factorization grows polynomially in $q$ at fixed $n$; snmmAGQ, by contrast, integrates over a product quadrature grid whose number of nodes grows exponentially in $q$.

To assess robustness to misspecified model formulations, Online Appendix B repeats the sine-curve study, with the error distribution and the random-effect distribution replaced by a $t$-distribution. Coverage is largely unaffected for all three methods in both cases. The main difference is in the number of failed fits, with the failure rates increasing under heavy-tailed random effects, reaching $8\%$ for assist.

As a second simulation example, we consider a bell-shaped curve, generated from the model
\begin{equation*}
y_{ij}
= 1 + b_{1i}
+ \exp\!\left( -\frac{1}{2}(t_{j} - b_{2i})^2 \right) + \varepsilon_{ij},
\quad i = 1, \ldots, n,\; j = 1, \ldots, m,
\end{equation*}
with $t_{j}$ taken at $m$ equally spaced points in $[-2,2]$, and with parameters $b_{1i}$ and $b_{2i}$ representing a subject vertical and horizontal shift respectively. We generate $\mathbf{b}_i \sim N(\mathbf{0}, \sigma^2 \mathbf{D})$ and
\begin{equation*}
\mathbf{D} = \begin{pmatrix} 2 & 1 \\ 1 & 2 \end{pmatrix}.
\end{equation*}
The off-diagonal entries induce a correlation of $\rho = 0.5$ between the two random effects. The correlation is introduced as an additional estimated parameter in the models. We consider parameter values: (1) $\sigma \in \{0.2, 0.4\}$; (2) $m \in \{10, 20\}$; (3) $n \in \{10, 50\}$; for a total of $8$ data settings, each simulated 500 times.

For this simulation, the knots for snmmTMB are placed with the knot scaling procedure described in Section \ref{sec: Knot Selection}: specifically, the knots are fixed in the interval $[0,1]$, and the internal function $\gamma$ is scaled accordingly to take values in that range. We specify the \texttt{assist} model with the space of linear functions going through the origin as null space, and the reproducing kernel of the cubic spline space on $[0,1]$ orthogonal to it, with an unstructured random-effects covariance. snmmAGQ uses $5^2 = 25$ quadrature nodes; all other specifications are as in the first simulation.

\begin{figure}
\centering
    \includegraphics[width=\linewidth]{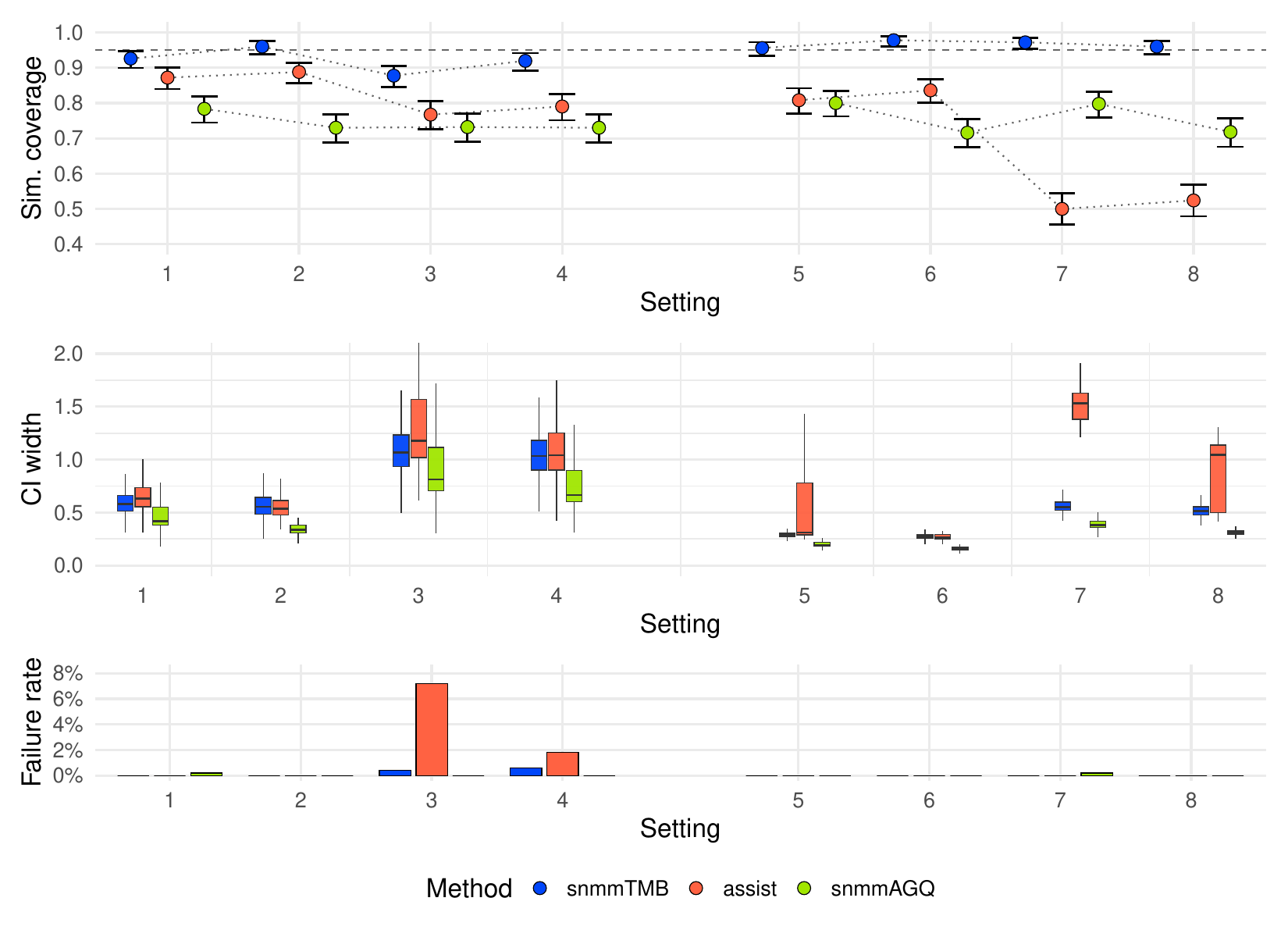}
    \caption{\textit{Results of the bell curve simulation on the population curve}. The first four settings correspond to $(n, m, \sigma)$ equal to $(10,10,0.2)$, $(10,20,0.2)$, $(10,10,0.4)$, and $(10,20,0.4)$. The remaining four follow the same order but with $n = 50$. The top panel displays the average simultaneous coverage of the population curve across 500 replications; bars indicate 95\% Clopper--Pearson binomial confidence intervals for the coverage probability. The middle panel shows boxplots of the confidence-band width. The bottom panel shows the percentage of replications that failed to produce a fit.}
    \label{fig:bellcurve}
\end{figure}

Figure~\ref{fig:bellcurve} shows the comparison. We observe that snmmTMB is still able to achieve coverage close to the nominal level. \texttt{assist} exhibits substantially lower coverage while producing, on average, wider confidence bands, and its coverage deteriorates as the number of subjects grows. This behavior appears to be related to difficulties in selecting an appropriate level of smoothness, with a tendency toward undersmoothing that leads to undercoverage; we present a more in depth analysis for a single setting of the grid in Online Appendix C. Finally, snmmAGQ undercovers throughout, consistent with the first simulation. The overall average number of basis functions selected by AIC among all settings for snmmAGQ is $6.1$.

\section{Application}\label{sec:Application}

We illustrate our approach using a longitudinal subsample of the Social Medical Survey of Children Attending Child Health Clinics (SMOCC), which followed Dutch children born in 1988--1989 with repeated measurements of height during the first two years of life \citep{Herngreen1994}. The original study investigated early childhood growth in relation to socioeconomic status; however, socioeconomic information is not included in the publicly available subsample of the data considered here. The
aim of this section is to demonstrate how longitudinal growth data of this type
can be modeled within the SNMM framework, to compare the results against another growth-curve model, and to describe a parametric bootstrap procedure that can be used in applied settings to assess the adequacy of the underlying
Laplace approximation.

After removing incomplete cases, the data contain $1904$ height measurements on
$200$ subjects, with between $5$ and $12$ measurements per subject (median
$10$). We include two covariates: sex (coded $0$ female, $1$ male) and
gestational age (GA), measured as deviation in weeks from the average pregnancy
length of $40$ weeks. Several candidate model specifications were considered,
and the fixed-effects structure was chosen using the marginal Akaike information
criterion \citep{Vaida2005}. A broader discussion of model selection, including considerations for random effects, is provided in Section \ref{sec:Discussion}. 

The following model was fit to the data:
\begin{equation}
\mathrm{hgt}_{ij} = \beta_0 + \beta_1 \mathrm{sex}_i + b_{1i}
  + \exp(\beta_2 \mathrm{sex}_i)\,
    f\!\left(\mathrm{age}_{ij} + \beta_3 \mathrm{GA}_i + b_{2i}\right)
  + \epsilon_{ij},
\label{eq:smocc}
\end{equation}
with
\begin{equation*}
\begin{pmatrix} b_{1i} \\ b_{2i} \end{pmatrix} \sim
N\!\left(\mathbf{0}, \Sigma\right), \qquad
\Sigma = \begin{pmatrix}
  \sigma_{b_1}^2 & \rho\,\sigma_{b_1}\sigma_{b_2} \\
  \rho\,\sigma_{b_1}\sigma_{b_2} & \sigma_{b_2}^2
\end{pmatrix}, \qquad
\epsilon_i \sim N(0, \sigma^2 I).
\end{equation*}
The correlation $\rho$ is parameterised as $\rho = \tanh(\tilde\rho)$, with $\tilde\rho$ an unconstrained parameter.

The response (height in centimeters) is modeled with a penalized cubic B-spline
basis of $K = 15$ functions, subject to sum-to-zero constraint. The fixed coefficients represent: $\beta_0$, the baseline height level for female subjects (sex $=0$); $\beta_1$, the intercept difference between males and females; $\beta_2$, a sex-specific scaling of the population growth trajectory, which accounts for differences in growth rate; $\beta_3$, the shift along the age axis associated with gestational age, so that differences in weeks of prematurity or post-term birth translate into horizontal shifts of the growth curve. The random effect $b_{1i}$ accounts for subject-specific deviations in overall height level, while $b_{2i}$ allows individual-specific timing shifts of the growth trajectory relative to the population curve.

We compare the results of \texttt{snmmTMB} with those of the SITAR model \citep{Cole2010}, implemented in the R package \texttt{sitar}, developed specifically for this type of growth-curve data setting. SITAR represents the mean growth curve by a natural cubic spline and allows
subjects to differ from it through up to three effects termed \textit{size} (a vertical shift),
\textit{tempo} (a horizontal shift), and \textit{velocity} (a scaling of the time axis).
Covariates enter linearly on these transformation parameters, and the model is
estimated with \texttt{nlme}. We use \texttt{sitar} to fit a model as close as
possible to that of \eqref{eq:smocc}, excluding only $\beta_2$, which has no
counterpart in the \texttt{sitar} formulation and cannot be included. The
degrees of freedom of the spline need to be specified in advance; we chose the
value by minimizing BIC over the set $\{3, \ldots, 20\}$, obtaining $df = 5$.

Table~\ref{tab:smocc_param_estimates} shows that the two methods agree closely. The sex effect on the intercept, $\beta_1$, indicates that male subjects are on average about $1.6$~cm taller at birth than female subjects. The sex-specific scaling of the growth trajectory, $\beta_2$, is not significant, giving no evidence that growth amplitude differs between sexes. The gestational-age effect on the time shift, $\beta_3$, is close to $-1$, so that a week of prematurity or post-maturity translates almost one-to-one into a corresponding shift along the age axis of the growth curve. The two random effects are strongly negatively correlated ($\hat\rho = -0.86$, 95\% CI $(-0.89, -0.81)$); this suggests that the vertical and horizontal shifts are, to a large extent, interchangeable ways of encoding the same feature of a subject's trajectory: a child who is taller throughout can be represented either by a larger $b_{1i}$ or by a $b_{2i}$ that places them further along the population curve. Finally, the estimated intercept $\beta_0$ differs between methods only because of definition: \texttt{snmmTMB}'s $\beta_0$ represents the intercept added to a sum-to-zero-constrained spline, while \texttt{sitar}'s intercept is by construction the height of the mean curve at the leftmost knot.

\begin{table}[!htbp]
\centering
\begin{tabular}{lcccc}
\hline
 & \multicolumn{2}{c}{snmmTMB} & \multicolumn{2}{c}{\texttt{sitar}} \\
\cline{2-3}\cline{4-5}
Quantity & Estimate & 95\% CI & Estimate & 95\% CI \\
\hline
Intercept $\beta_0$        & 72.89 & (72.40, 73.39) & 46.44 & (46.08, 46.80) \\
Sex (male) intercept $\beta_1$               &  1.67 & (1.12, 2.22)   &  1.65 & (1.17, 2.12)   \\
Sex (male) scale $\beta_2$            &  0.00 & (-0.02, 0.02)  & ---    & ---              \\
GA, time shift $\beta_3$  &  0.96 & (-1.12, -0.81)   &  0.95 & (-1.10, -0.79)   \\
$\sigma_{b_1}$                         &  3.16 & (2.85, 3.51)   &  3.12 & (2.81, 3.46)   \\
$\sigma_{b_2}$                         &  3.79 & (3.37, 4.27)   &  3.64 & (3.23, 4.11)   \\
$\rho$                   & -0.86 & (-0.89, -0.81) & -0.85 & (-0.89, -0.80) \\
$\sigma$                               &  1.04 & (1.00, 1.08)   &  1.05 & (1.01, 1.09)   \\
Spline df              & 10.2 & ---              &  5     & ---              \\
\hline
\end{tabular}
\caption{Parameter estimates for the SMOCC data under model \eqref{eq:smocc} and
under the corresponding \texttt{sitar} fit, with $95\%$ Wald intervals.
Intervals for standard deviations are computed on the log scale and
back-transformed; the interval for $\rho$ is computed for $\tilde\rho$ and
back-transformed. The snmmTMB spline degrees of freedom are the effective
degrees of freedom of the penalized fit; the
\texttt{sitar} value is the fixed (unpenalized) degrees of freedom chosen by
BIC.}
\label{tab:smocc_param_estimates}
\end{table}

\begin{figure}
\vspace{-1.1cm}
    \centering
    \includegraphics[width=0.9\linewidth]{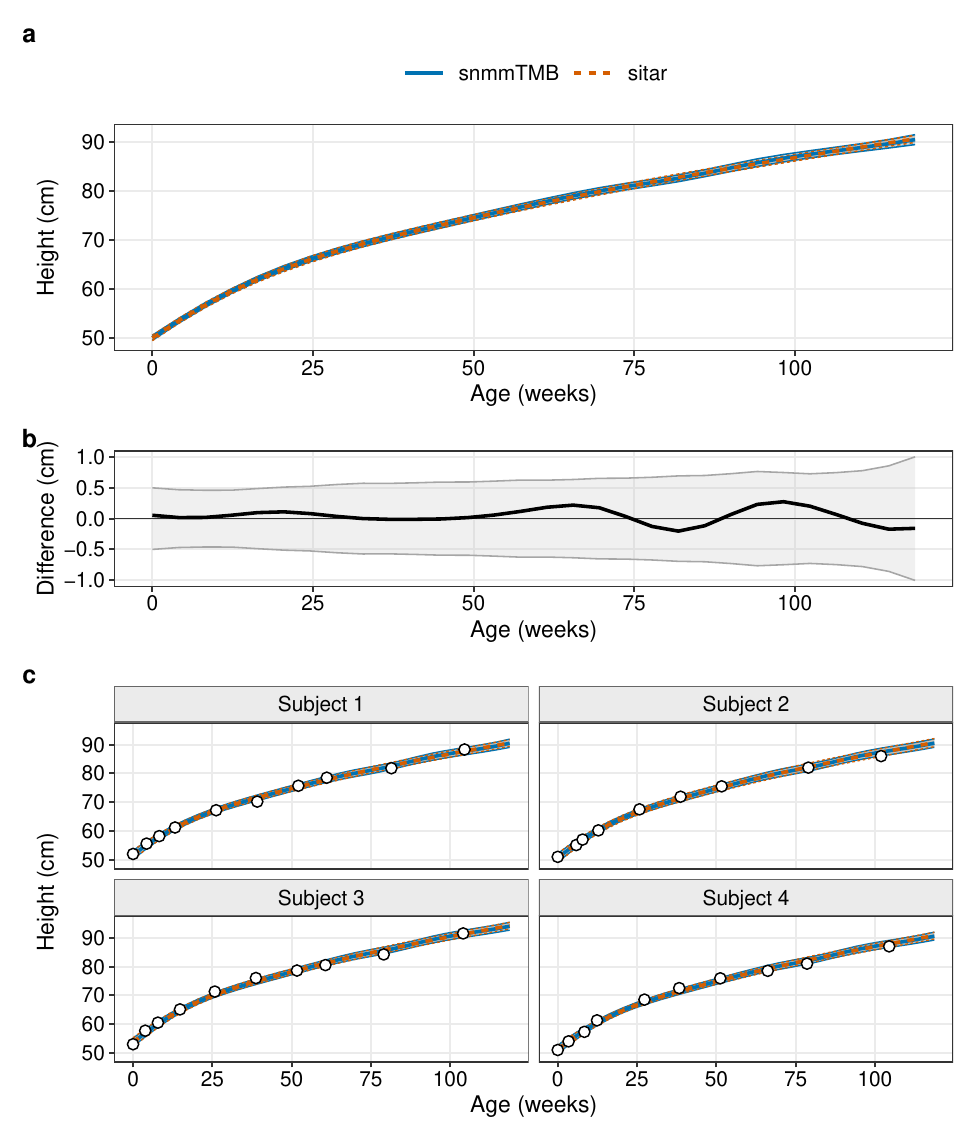}
    \caption{\textit{Population and subject-specific growth curves for the SMOCC application with snmmTMB and sitar.} (a) Population curves at the reference level (sex = 0, gestational age = 40 weeks, no random effects), with simultaneous 95\% bands. (b) Difference between the two fitted population curves (snmmTMB $-$ sitar); the shaded band is the snmmTMB simultaneous 95\% band. (c) Subject-specific fitted curves for four subjects, evaluated at each subject's own covariates and estimated random effects, with observed data as circles.}
    \label{fig:smocc_curves}
\end{figure}

Figure~\ref{fig:smocc_curves} presents the resulting population and subject curves, which are likewise almost indistinguishable between the two methods. Unlike for \texttt{snmmTMB}, a confidence band is not directly available for the \texttt{sitar} trajectory and is instead obtained by bootstrapping ($200$ samples). The estimated trajectory exhibits the well-known pattern of rapid increase during the first six months \citep{WHOgrowth}, followed by more gradual growth over the remainder of the observation period. Between-subject variability appears limited, with individual curves closely tracking the population trend.

Neither method is computationally demanding for this sample size, but \texttt{sitar} is substantially faster to fit ($1.1$s vs.\ $64.2$s); if confidence intervals for the fitted curve are needed, however, \texttt{sitar} requires an additional bootstrap step, which raises its total runtime.

\subsection{Accuracy of the Laplace approximation}

For the Wald confidence intervals reported in Table~\ref{tab:smocc_param_estimates} to be reliable, the marginal log-likelihood should be well approximated by a quadratic function in a neighborhood of its maximum. A parametric bootstrap can be used to assess this assumption and to investigate potential bias in the parameter estimates and their associated standard errors \citep[Section~4.2]{sorensenLongitudinalModelingAgeDependent2023}. In this procedure, multiple datasets are generated from the fitted model by resampling the subject-specific random effects and noise, and the model is re-estimated for each simulated dataset. We note that the random effects associated with the spline coefficients serve only as a computational device for representing the smooth function: consequently, in generating the bootstrap samples, we use the fitted parameter values directly rather than drawing a new smooth function for each replicate.
If the quadratic approximation is adequate, the empirical variability of the bootstrap estimates should be consistent with the standard errors obtained from the asymptotic covariance matrix.

\begin{figure}
\centering
    \includegraphics[width=1\linewidth]{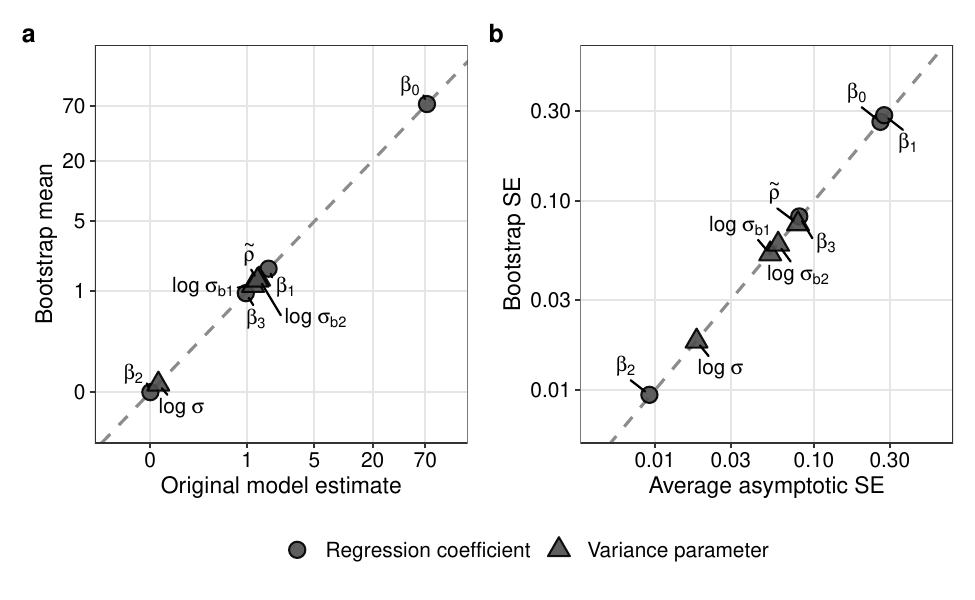}
    \caption{\textit{Parametric bootstrap assessment of bias and standard errors.} Circles are the fixed regression coefficients, triangles are the variance/correlation parameters ($\operatorname{log} \sigma_{b_1}$, $\operatorname{log} \sigma_{b_2}$, $\tilde\rho$, $\operatorname{log} \sigma$). Results from $500$ simulated datasets. Axes are log-scaled. (a) Original parameter estimates against their bootstrap averages. (b) Empirical standard deviation of the bootstrap estimates versus the average asymptotic standard errors.}
    \label{fig: smocc_parametric_bootstrap}
\end{figure}

Figure~\ref{fig: smocc_parametric_bootstrap} summarizes the results from $500$ bootstrap replicates by comparing the original parameter estimates with their bootstrap means, and the empirical standard deviations of the bootstrap estimates with the corresponding average asymptotic standard errors.
Overall, the bootstrap means closely track the original parameter estimates, suggesting little evidence of bias, while the asymptotic standard errors provide a good approximation to the empirical sampling variability.
The same bootstrap-based procedure could alternatively be used to construct confidence intervals directly, although computing the extreme quantiles needed to determine interval endpoints in a stable way would require a larger number of replicates.

\section{Discussion}\label{sec:Discussion}

We have presented a procedure for estimation and inference for the semiparametric nonlinear mixed-effects model of \citet{Ke2001}. We introduce two main adaptations: (i) the population trajectory is represented through a penalized spline expressed in mixed-model form, which allows us to estimate the smoothing parameter as a variance component, without requiring additional smoothness selection procedures; (ii) the marginal likelihood is evaluated through the Laplace approximation, with the required derivatives obtained by AD. Across simulations the procedure obtained accurate point estimates and improved coverage of the population curve, with narrower confidence bands, when compared to other estimation approaches. Application to infant height data between birth and two years produced population trajectories and parameter estimates consistent with the established pattern of early-life growth, and almost identical to the growth curve method \texttt{sitar}, illustrating the practical applicability of the proposed framework.

An alternative estimation strategy for models of this kind is the EM algorithm \citep{Dempster1977}, treating random effects and spline coefficients as missing data. Stochastic-approximation variants of EM (SAEM) have been used for related forms of SNMMs \citep{ArribasGil2013, Mrquez2023}, but only in settings where the nonlinear function does not depend on the random effects. Moreover, EM does not directly provide standard errors, requiring additional computational steps \citep{Louis1982}. A Bayesian implementation, for instance using Stan \citep{Carpenter2017}, is another possible option, and could yield full posterior inference for both  spline coefficients and fixed effects.

Model selection for the application example was performed on fixed-effects using marginal AIC, while keeping the random-effects structure fixed. Selecting random effects would instead require using a conditional AIC \citep{Vaida2005}, but the standard form does not account for uncertainty in estimated variance parameters and therefore tends to favor inclusion of any random effects with positive variance \citep{Greven2010}. An adjusted conditional AIC that incorporates variance and smoothing-parameter uncertainty has been proposed for general smooth models by \citet{Wood2016}; extending this approach to the SNMM framework represents a useful direction for future work.

Identifiability is an important aspect to consider when fitting an SNMM. Different combinations of spline components and model parameters can yield equivalent likelihoods, for example when shifts or scaling of the population curve are confounded with intercept or scale parameters. Although these issues are highly case-specific, they can often be addressed by imposing appropriate constraints on the spline, such as a sum-to-zero constraint to ensure identifiability of an intercept term, or by restricting scale parameters to be positive.
The fit of an SNMM can also shift horizontally through the transformation $\gamma$. In that case, one way to identify the curve is to anchor it at a specific point, for instance by fixing $f$ to a chosen value at a reference argument, which removes the freedom to translate $f$ along its axis. We impose no such constraint, and rely instead on the mean of the shift random effect being fixed at zero, with no separate population-level shift term, so that the shifts represent random fluctuations around an identified mean curve. A population-level shift can still be included when it multiplies an observed covariate, as with the gestational-age term $\beta_3 \cdot \mathrm{GA}_i$ in the SMOCC application of Section~\ref{sec:Application}: a translation of $f$ by a constant $-c$ can only be reabsorbed by adding $c$ to every subject's shift, whereas $\beta_3 \cdot \mathrm{GA}_i$ varies across $i$ with $\mathrm{GA}_i$, so no such translation leaves the fit unchanged. Theoretical discussions of identifiability of similar models can be found in \citet{Kneip1988} and \citet{Lindstrom1995}.

We used a B-spline basis to represent the population curve. Other spline bases could be considered as well, provided that their evaluation and derivatives can be implemented in a differentiable manner in \texttt{C++}, to ensure compatibility with AD.

In many applications, it is desirable to enforce shape constraints on the population curve (e.g. monotonicity, convexity) based on prior knowledge of the underlying process, as unconstrained models might give implausible results. These constraints are often imposed by reparametrizing the spline coefficients \citep{Pya2014}; however, this approach is hard to apply in our framework, as the coefficients are represented as mixed effects to enforce smoothness. A more accessible alternative is to encourage shape properties through penalties on the spline derivatives. This strategy is discussed further in Online Appendix D.

The implementation in \texttt{TMB} is naturally flexible and can accommodate several extensions to the model specification that were not explored here. These include relaxing the Gaussian error assumption to allow non-Gaussian responses through appropriate link functions and adopting alternative random-effects covariance structures. Incorporating multiple smooth components or interaction terms through additional penalized splines could also be possible, although in such settings careful attention to identifiability and appropriate constraints becomes even more important.

\section*{Disclosure statement}\label{disclosure-statement}

The authors declare no conflict of interest.

\section*{Data Availability Statement}\label{data-availability-statement}


The code for reproducing the simulated and real examples and the SMOCC dataset used as example in the article are available as Supplementary Material and on GitHub at \url{https://github.com/matteodales/snmmTMB}.
\vspace{0.5cm}
\bigskip
\phantomsection\label{supplementary-material}

\begin{center}

{\large\bf SUPPLEMENTARY MATERIAL}

\end{center}

\begin{description}

\item[Online Appendices:]
Appendices discussing: (A) simultaneous confidence band construction for \texttt{assist}, (B) misspecified variants of the sine curve simulation, (C) comparison of effective degrees of freedom, (D) monotonicity penalty (.pdf file).

\item[snmmTMB code:]
Code for reproducing the simulated and real examples, containing the SMOCC dataset used as example in the article. (zip file).

\end{description}

\bibliography{sample.bib}

\end{document}


\maketitle

\renewcommand{\thesection}{\Alph{section}}
\renewcommand{\figurename}{Supplementary Figure}

\section{Simultaneous confidence bands for \texttt{assist}}
In the simulation studies of Section~3, we are interested in understanding if the methods are able to recover the shape of the population spline as a whole. In order to do that, we compare them on coverage of the entire curve, rather than at single points. This requires a simultaneous confidence band; however the \texttt{R} package \texttt{assist} only reports approximate Bayesian confidence intervals, which hold an across-the-function coverage property \parencite{Wahba1983}. We describe here how we instead construct a simultaneous band, following the same method we use for \texttt{snmmTMB} \parencite{sorensenLongitudinalModelingAgeDependent2023}.
 
Consider as an example the sine-curve simulation, written as
\begin{equation*}
  y_{ij} = \beta_0 + b_{1i} + e^{b_{2i}} f(u_{ij}) + \varepsilon_{ij}, \qquad u_{ij} = t_{ij} - \frac{e^{b_{3i}}}{1 + e^{b_{3i}}}, \qquad \varepsilon_{ij} \sim N(0, \sigma^2),
\end{equation*}
with $\mathbf{b}_i = (b_{1i}, b_{2i}, b_{3i})^\top \sim N(\mathbf{0}, \sigma^2 \mathbf{D})$. The target of inference is the population curve, obtained by setting $\mathbf{b} = \mathbf{0}$ and evaluated on a reporting grid $x_1, \dots, x_G$; we seek a band $\hat f_k \pm \tilde c_{1-\alpha} \sqrt{\hat v_k}$ whose coverage is $1-\alpha$ simultaneously over the grid.
 
The construction requires a matrix $\mathbf{X}_{\mathrm{new}}$ evaluating the basis on the grid, and the covariance matrix $\operatorname{Cov}(\hat{\boldsymbol{\beta}})$ of the spline coefficients. Given these,
\begin{equation}
  \hat{\mathbf{f}} = \mathbf{X}_{\mathrm{new}} \hat{\boldsymbol{\beta}}, \qquad \hat{\mathbf{v}} = \operatorname{diag}\!\big( \mathbf{X}_{\mathrm{new}} \operatorname{Cov}(\hat{\boldsymbol{\beta}}) \mathbf{X}_{\mathrm{new}}^\top \big),
  \label{eq:sup-fv}
\end{equation}
and the simultaneous critical value $\tilde c_{1-\alpha}$ is the $(1-\alpha)$ quantile of
\begin{equation}
  r = \max_k \frac{\big| \mathbf{X}_{\mathrm{new}} (\boldsymbol{\beta}^{*} - \hat{\boldsymbol{\beta}}) \big|_k}{\sqrt{\hat v_k}}, \qquad (\boldsymbol{\beta}^{*} - \hat{\boldsymbol{\beta}}) \sim N\!\big(\mathbf{0}, \operatorname{Cov}(\hat{\boldsymbol{\beta}})\big),
  \label{eq:sup-r}
\end{equation}
obtained by simulation.\\
 
To make use of this construction, we need to express the \texttt{assist} fit in the form $\mathbf{X}\boldsymbol{\beta}$. Following \cite{Ke2001}, the method works in a reproducing kernel Hilbert space decomposed as $\mathcal{H} = \mathcal{H}_0 \oplus \mathcal{H}_1$, where $\mathcal{H}_0 = \operatorname{span}\{\varphi_1, \dots, \varphi_M\}$ is an unpenalized null space and $\mathcal{H}_1$ is represented by a reproducing kernel $R_1(\cdot,\cdot)$; the penalty is the squared norm of the projection of $f$ onto $\mathcal{H}_1$. For example, in the sine-curve simulation the null space is the space of periodic functions $\operatorname{span}\{\sin 2\pi u, \cos 2\pi u\}$ and $R_1$ is the corresponding periodic kernel. The fitted function is then a combination of the null space functions and one representer per observation,
\begin{equation}
  \hat f(x) = \sum_{\nu = 1}^{M} d_\nu \varphi_\nu(x) \; + \sum_{(ij)} c_{ij} R_1\!\big(x, \hat u_{ij}\big),
  \label{eq:sup-representer}
\end{equation}
where the $\hat u_{ij}$ are the observed time points shifted by the estimated horizontal random effect, $\mathbf{d}$ collects the null space coefficients and $\mathbf{c}$ the representer coefficients.
 
Eq. (\ref{eq:sup-representer}) is linear in the coefficients, so collecting $\boldsymbol{\beta} = (\beta_0, \mathbf{d}^\top, \mathbf{c}^\top)^\top \in \mathbb{R}^{1+M+N}$ and defining the grid matrices $\boldsymbol{\Phi}_{k\nu} = \varphi_\nu(x_k)$ and $\boldsymbol{\Sigma}_{k,(ij)} = R_1(x_k, \hat u_{ij})$ gives
\begin{equation*}
  \mathbf{X}_{\mathrm{new}} = \big[\ \mathbf{1} \quad \boldsymbol{\Phi} \quad \boldsymbol{\Sigma}\ \big] \in \mathbb{R}^{G \times (1+M+N)},
\end{equation*}
which has the required form $\hat f = \mathbf{X}_{\mathrm{new}}\hat{\boldsymbol{\beta}}$. 
 
For the covariance of the coefficients we use the standard smoother expression
\begin{equation*}
  \operatorname{Cov}(\hat{\boldsymbol{\beta}}) = \hat\sigma^2 \big( \mathbf{X}^\top \boldsymbol{\Lambda}^{-1} \mathbf{X} + \hat\lambda\, \mathbf{S} \big)^{-1},
\end{equation*}
where $\mathbf{X}$ is the design matrix of observed data, $\hat\lambda$ is the estimated smoothing parameter, $\mathbf{S}$ the penalty matrix, and $\boldsymbol{\Lambda}$ the scaled working covariance of the response, $\operatorname{Var}(\mathbf{y}) = \hat\sigma^2 \boldsymbol{\Lambda}$.
 
Substituting these quantities in Equations (\ref{eq:sup-fv})-(\ref{eq:sup-r}) yields the simultaneous band reported in Section~3; the implementation is available in the supplementary code.

\section{Non-Gaussian variants of the sine curve simulation}
\label{app:nongauss}
 
The simulations of Section~3 generate data from well-specified models. Here, we test robustness to breaking of the Gaussian assumption for the residuals and the random effects in the sine curve simulation. The grid of $16$ settings, the number of replications and the configuration of the three methods remain the same, so that the results are directly comparable with Figure~1.
 
In the first variant the residuals are drawn as $\varepsilon_{ij} = \sigma\, T_{ij} / \sqrt{\nu/(\nu-2)}$, with $T_{ij} \sim t_\nu$ and $\nu = 3$, so that $\operatorname{Var}(\varepsilon_{ij}) = \sigma^2$. In the second variant the residuals remain Gaussian and the random effects are drawn from an elliptical multivariate $t$ with $\nu = 3$, rescaled so that $\operatorname{Cov}(\boldsymbol{b}_i) = \sigma^2 \mathbf{D}$ is preserved. In both cases the first two moments of the data-generating process coincide with those of the Gaussian model, and the misspecification is confined to the distribution shape.
 
The results are reported in Supplementary Figures~\ref{suppfig:nongauss_error} and~\ref{suppfig:nongauss_randomeff}. Coverage is comparable to the Gaussian case for all three methods in nearly every setting. Confidence-band widths are generally more variable, with visibly wider boxplots. The two variants mainly differ in the proportion of failed fits: under heavy-tailed random effects, the failure rates of \texttt{assist} rise to $2.1\%$ on average, and to $8.2\%$ in one setting with $n = 10$. This is likely due to heavy-tailed random effects occasionally producing extreme subjects, which increase the number of failures.

\begin{figure}
\centering
    \includegraphics[width=\linewidth]{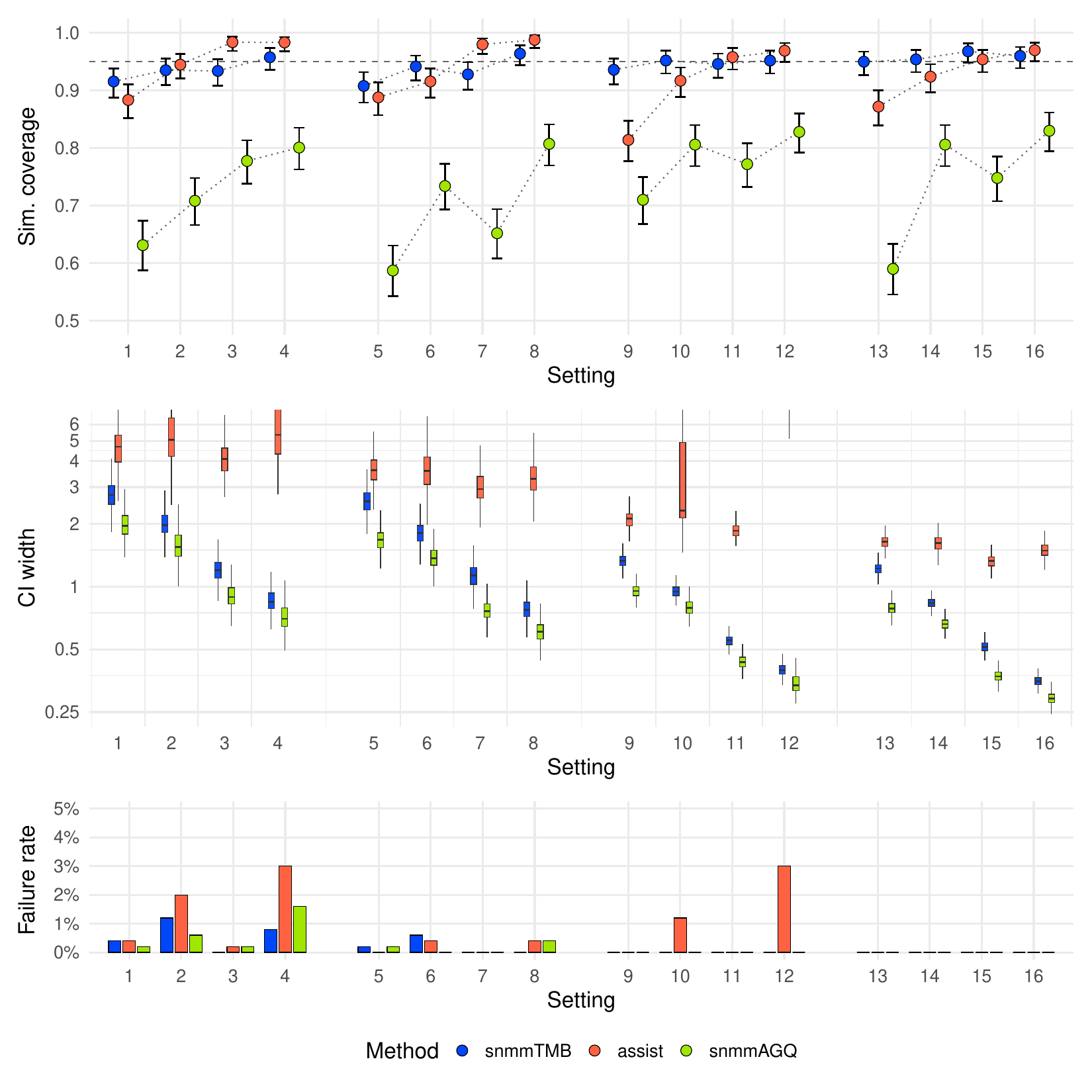}
    \caption{\textit{Results of the sine curve simulation on the population curve, with $t_3$ residuals}. The residuals are drawn from a $t$ distribution with $3$ degrees of freedom, rescaled to variance $\sigma^2$; the random effects remain Gaussian. Settings, panels and number of replications are as in Figure 1.}
    \label{suppfig:nongauss_error}
\end{figure}

\begin{figure}
\centering
    \includegraphics[width=\linewidth]{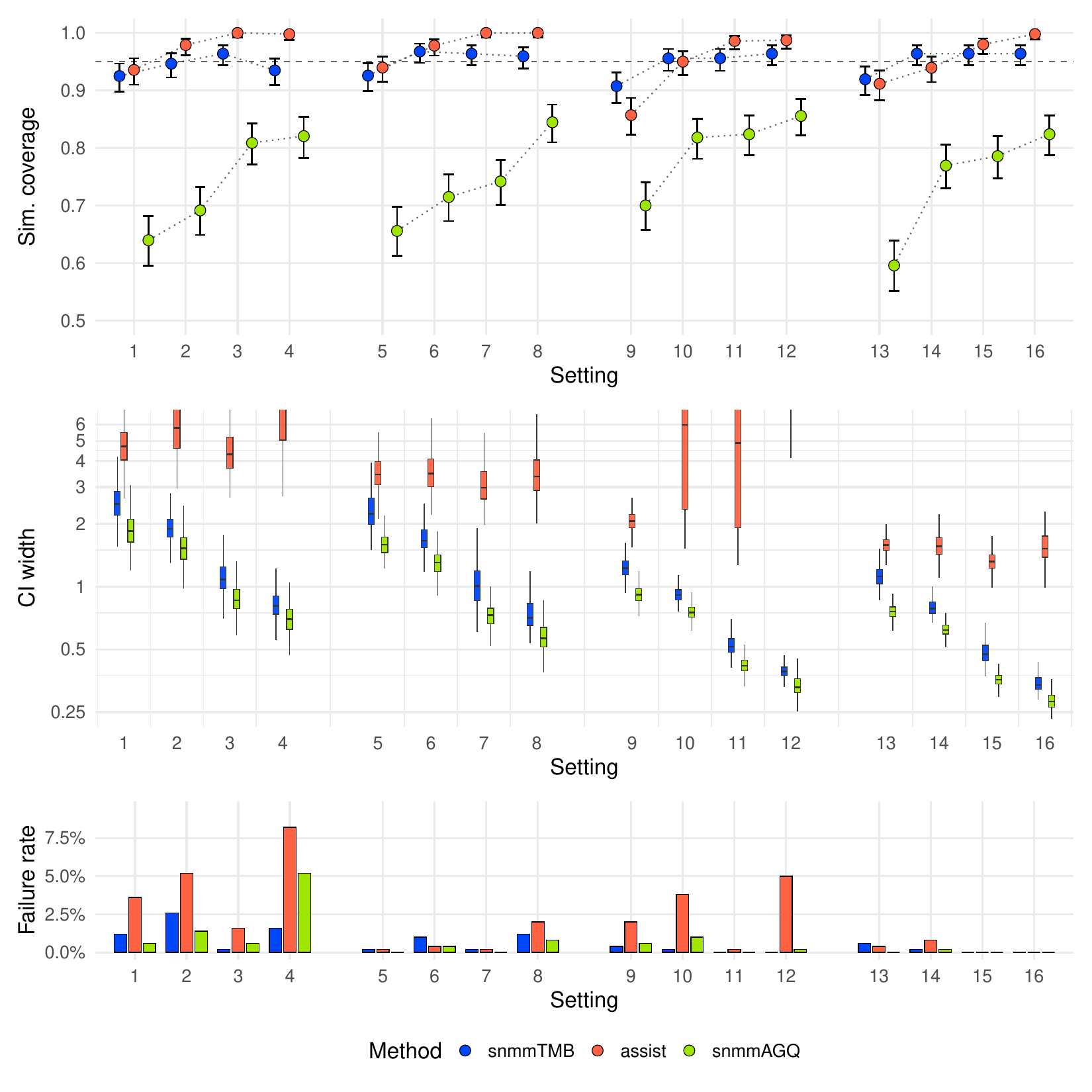}
    \caption{\textit{Results of the sine curve simulation on the population curve, with $t_3$ random effects}. The random effects are drawn from an elliptical multivariate $t$ with $3$ degrees of freedom, rescaled to covariance $\sigma^2 \boldsymbol{D}$; the residuals remain Gaussian. Settings, panels and number of replications are as in Figure 1.}
    \label{suppfig:nongauss_randomeff}
\end{figure}

\section{EDF comparison}
The coverage of \texttt{assist} in the bell curve simulation of Section 3 deteriorates in settings with higher variability, especially with a larger number of subjects. Although we do not have a definitive explanation for this behaviour, we have observed that in some replications the curve estimated by \texttt{assist} is very undersmoothed and wiggly. To illustrate this, we compare the effective degrees of freedom of the estimated splines for the three methods on a single setting of the bell curve grid, with $n = 10$, $m = 10$ and $\sigma = 0.4$, over $50$ replications.

For snmmTMB this quantity is not returned by the fitting routine. We therefore compute it as in \cite[Section 5.4.2]{Wood2017}, which defines the effective degrees of freedom of an ordinary penalized spline as
\begin{equation*}
\mathrm{EDF} = \operatorname{tr}\left[(\mathbf{X}^\top \mathbf{X} + \lambda \mathbf{S})^{-1} \mathbf{X}^\top \mathbf{X}\right],
\end{equation*}
where $\mathbf{X}$ is the spline design matrix and $\mathbf{S}$ the penalty matrix. For \texttt{assist} the effective degrees of freedom can be extracted directly from the fitted object in \texttt{R}. For snmmAGQ the shape spline is unpenalized, so its EDF is the rank of the
centered B-spline design matrix, that is the number of basis functions minus one for the sum-to-zero constraint. Since the number of basis functions is selected by AIC for every dataset (between 4 and 10, Section 3), this value varies across replications, taking integer values between 3 and 9.

\begin{figure}
\centering
    \includegraphics[width=\linewidth]{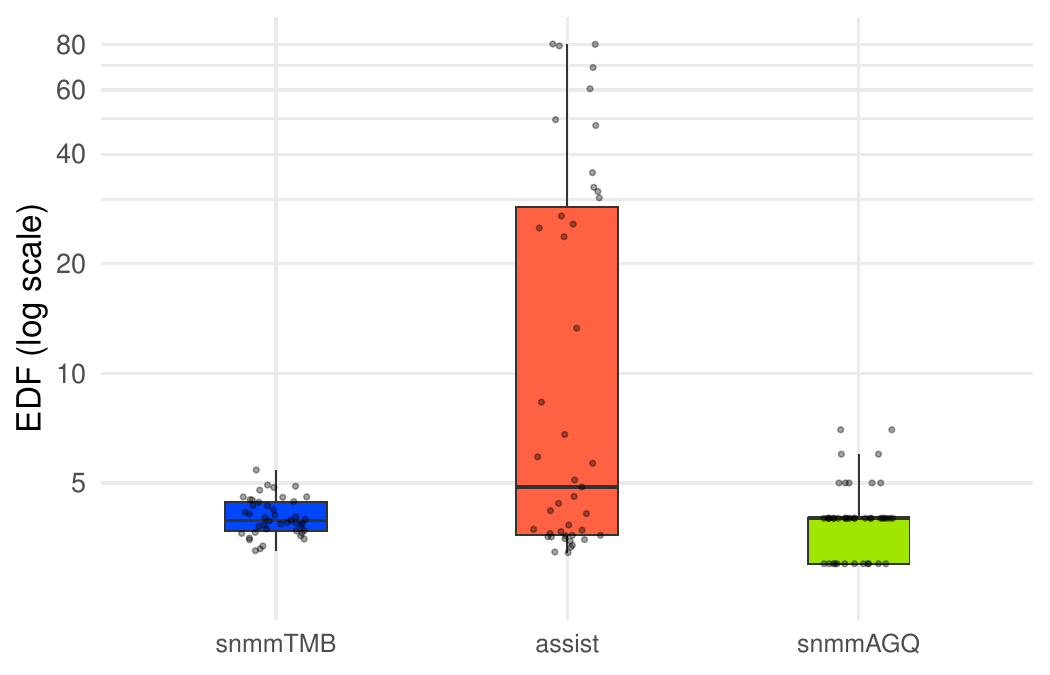}
    \caption{\textit{Effective degrees of freedom of the estimated splines in the bell curve simulation}. Boxplots over $50$ replications of the bell curve simulation with $n = 10$, $m = 10$ and $\sigma = 0.4$.}
    \label{suppfig:edf}
\end{figure}

Supplementary Figure~\ref{suppfig:edf} shows the resulting EDFs. The median value for \texttt{assist} is close to that of snmmTMB but, as expected, we see an instability in the selection of the level of smoothness, with some iterations approaching a pure interpolation of the data. The EDF of the estimated curve for snmmTMB are instead very concentrated around their mean value $4.06$.

\section{Monotonicity}

As mentioned in the Discussion, it is often desirable to impose shape constraints such as monotonicity when prior knowledge about the underlying population trajectory is available. In our framework, direct constraints on spline coefficients are difficult to impose because smoothness is enforced by representing penalized components as random effects within a mixed-model formulation. Instead, we suggest a penalty-based approach that promotes monotonicity while preserving the mixed-model structure and remaining compatible with automatic differentiation. Similar asymmetric penalties have been proposed in \cite{Eilers2005, Bollaerts2006, Hofner2014}.

Through the TMB implementation, additional penalty terms can easily be incorporated directly into the negative log-likelihood without altering the underlying model specification. Moreover, derivatives of B-splines can be evaluated efficiently using standard recursion formulas.

Let the population smooth be represented using a B-spline basis as
$$
f(x) = \sum_{k=1}^{K} \theta_k B_k(x),
$$
where $B_k(\cdot)$ are B-spline basis functions and $\theta_k$ are the corresponding coefficients. Under the mixed-model representation, the full vector of spline coefficients $\boldsymbol{\theta}$ can be reconstructed from the fixed and random components at each iteration of the optimization.
The derivative of a B-spline of degree $p$ can be expressed as a linear combination of two B-splines of degree $p-1$, which allows the derivative of the smooth to be written as
$$
f'(x) = \sum_{k=1}^{K} \theta_k B_k'(x),
$$
and evaluated at arbitrary points with a very small computational cost. Specifically, we obtain $B_k'$ from function \texttt{splineDesign} from R package \texttt{splines}. \\

To promote monotonicity, we can then evaluate $f'(x)$ on a dense grid of points $\{x_i\}_{i=1}^{M}$ spanning the domain of interest and introduce an asymmetric penalty that penalizes negative derivatives while leaving positive derivatives unaffected. Specifically, we define the penalty

$$
\lambda_{c} \sum_{i=1}^{M}
\left[ \min\big(f'(x_i),\,0\big) \right]^2,
$$

where $\lambda_{c} > 0$ controls the strength of the constraint. Since the function $\min(f'(x),0)$ is not differentiable, a smooth approximation is required for compatibility with automatic differentiation. We can replace it with the smooth approximation
$$
\min(f'(x),0) \approx \frac{1}{2} \left(\sqrt{f'(x)^2 + \varepsilon} - f'(x)\right),
$$
where $\varepsilon$ is a small positive constant that keeps the approximation differentiable at $f'(x) = 0$. We used $\varepsilon = 10^{-8}$ in this example, and tested automatic differentiation stability with $\varepsilon$ values in the range of $10^{-4}$ to $10^{-16}$, and observed no relevant difference. Only $\varepsilon = 0$ produced a failure, as expected. Using this approximation, for sufficiently large values of $\lambda_{c}$, the estimated curve is effectively constrained to be monotone increasing.

To illustrate the effect of the penalty, we present a simulation example in Supplementary Figure~\ref{suppfig: monotone_penalty}, which shows fitted curves obtained under increasing values of $\lambda_c$. The data are generated on the interval $[-2,2]$ from the function $y = x^3 - x$, which is not monotone. Accordingly, fitting the spline without the monotonicity penalty yields a correctly non-monotone estimate. As the penalty weight increases, the estimated population curve becomes progressively less decreasing.

We note that similar penalties could be put on the second derivative of the curve to induce convexity.

\begin{figure}
\includegraphics[width=1.2\linewidth, center]{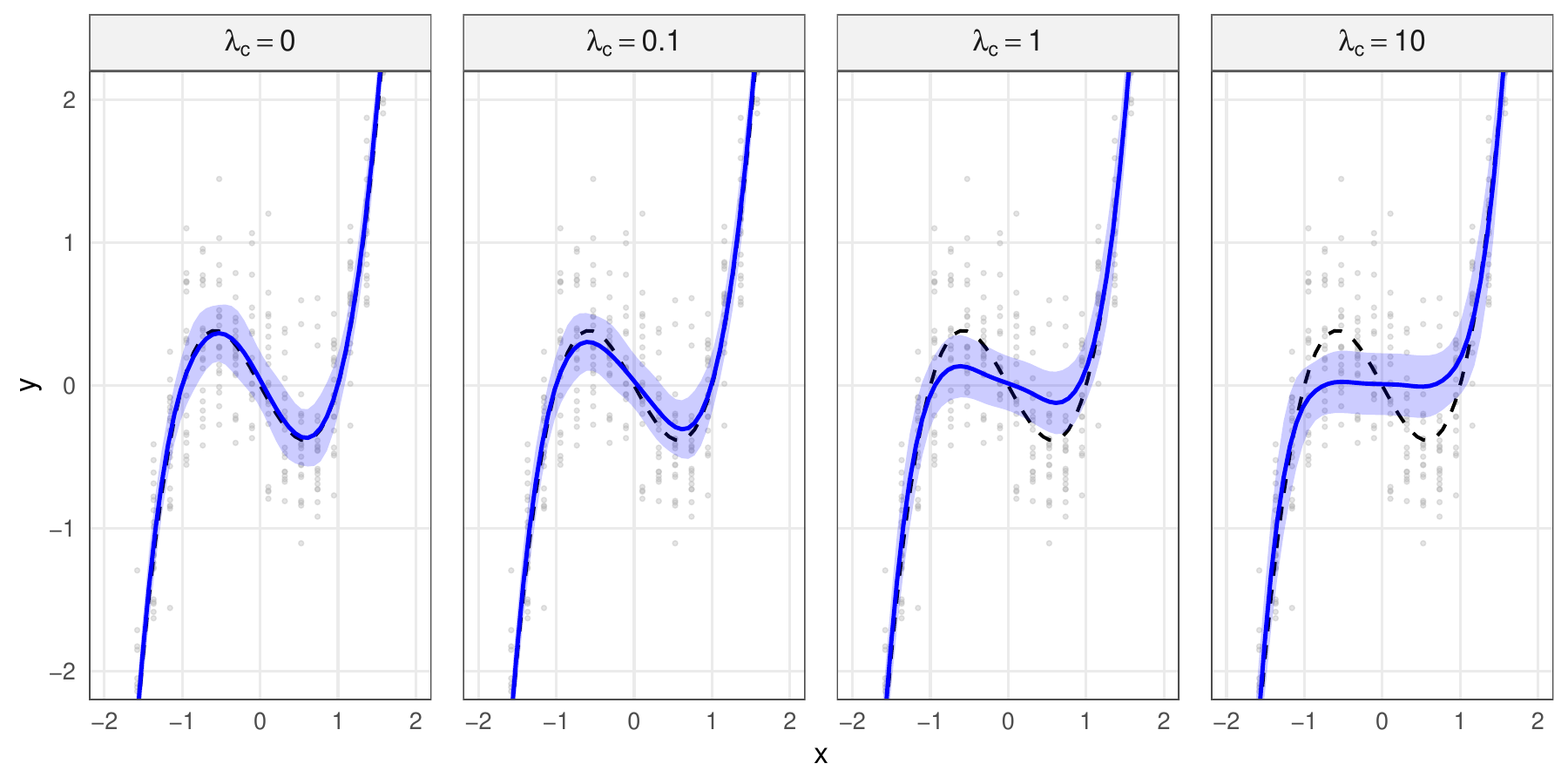}
\caption{\textit{Effect of the monotonicity penalty on spline estimation}. Simulated data are generated from the function $y = x^3 - x$. Points represent the observed data, the dashed line denotes the true function, and the solid line shows the estimated population curve with an increasing value of the monotonicity penalty $\lambda_c = 0, 0.1, 1, 10$, and its associated simultaneous confidence band. As $\lambda_c$ increases, the fitted curve becomes progressively less decreasing.}
\label{suppfig: monotone_penalty}
\end{figure}

\printbibliography